\documentclass[aps,onecolumn,11pt,floatfix,altaffilletter,tightenlines,showpacs,showkeys,notitlepage,nofootinbib]{revtex4-1}
\pdfoutput=1

\usepackage[colorlinks=true,citecolor=blue,linkcolor=blue]{hyperref}
\usepackage[normalem]{ulem}
\usepackage{amsmath,amssymb}
\usepackage{epsfig}
\usepackage{graphicx}
\usepackage{url}
\usepackage{color}
\usepackage{multirow}
\usepackage{placeins}
\usepackage[dvipsnames]{xcolor}
\usepackage{epstopdf}
\usepackage[abs]{overpic}

\usepackage{amssymb}
\usepackage{pifont}
\newcommand{\cmark}{\ding{51}}
\newcommand{\xmark}{\ding{55}}

\usepackage{tikz}
\usetikzlibrary{trees}
\usetikzlibrary{decorations.pathmorphing}
\usetikzlibrary{decorations.markings}

\definecolor{jblue}  {RGB}{20,50,100}
\definecolor{npurple}  {RGB} {153, 51, 204}
\definecolor{wred}   {RGB}{217,0,56}
\definecolor{white}   {RGB}{255,255,255}

\definecolor{korange}   {RGB}{235, 80,  43}
\definecolor{korange2}   {RGB}{245, 100,  63}
\definecolor{kyelloworange}   {RGB}{255, 210,  110}
\definecolor{kyelloworange2}   {RGB}{240, 170,  90}
\definecolor{kred}   {RGB}{204,  102, 153}
\definecolor{kpurple}   {RGB}{153,  61, 190}
\definecolor{kpurplelight}   {RGB}{213,  161, 230}

\usepackage{cleveref}
\usepackage{subfigure} 
\usepackage{lipsum}
\definecolor{red}{rgb}{1.0, 0, 0}
 \usepackage{gensymb}

\allowdisplaybreaks

\bibpunct{[}{]}{,}{n}{}{,}

\renewcommand{\vec}[1]{{\mathbf{#1}}}


\begin{document}

\title{Neutrino Astronomy with Supernova Neutrinos}

\author{Vedran Brdar}   \email{vbrdar@mpi-hd.mpg.de}
\author{Manfred Lindner}   \email{lindner@mpi-hd.mpg.de}
\author{Xun-Jie Xu} \email{xunjie.xu@mpi-hd.mpg.de}
\affiliation{Max-Planck-Institut f\"ur Kernphysik,
       69117~Heidelberg, Germany}

\begin{abstract}

Modern neutrino facilities will be able to detect a large number of neutrinos from the next Galactic supernova. We investigate the viability of the triangulation method to locate a core-collapse supernova by employing the neutrino arrival time differences at various detectors. We perform detailed numerical fits in order to determine the uncertainties of these time differences for the cases when the core collapses into a neutron star or a black hole. We provide a global picture by combining all the relevant current and future neutrino detectors. Our findings indicate that in the scenario of a neutron star formation, supernova can be located with precision of 1.5 and 3.5 degrees in declination and right ascension, respectively. For the black hole scenario, sub-degree precision can be reached.

\end{abstract}

\maketitle

\section{Introduction}
\label{sec:intro}
Core-collapse supernovae (SNe) are among the most energetic astrophysical events in the Universe and thus of great interest in astronomy, astrophysics as well as in
particle physics.
The recent SN1987A \cite{1987} event was observed in both optical and neutrino channels, which may be regarded as the dawn of  multi-messenger astronomy.
In the near future, more optical telescopes and neutrino detectors will be upgraded or constructed,
leading to potentially very robust observations of Galactic SNe\footnote{Recent simulations \cite{Andresen:2016pdt} indicate that SN explosions are likely to be highly asymmetric. SNe might, therefore, also be detected via gravitational waves (GW) by laser interferometers such as LIGO \cite{LIGO} and VIRGO \cite{TheLIGOScientific:2017qsa}.
}.
Observation of Galactic SNe would be especially important for the field of neutrino physics as it could resolve the long standing problem of neutrino mass ordering
\cite{Barger:2005it,Scholberg:2012id,Scholberg:2017czd}, unravel collective neutrino oscillation \cite{Duan:2010bg,Akhmedov:2017mcc}, severely constrain  electromagnetic properties of neutrinos \cite{Giunti:2014ixa}, reveal possible non-standard neutrino  interactions
\cite{Farzan:2002wx,Dent:2012mx,Harnik:2012ni,Das:2017iuj,Dighe:2017sur},
or provide a crucial information  for the identification of dark matter (DM) and its
interactions with visible matter \cite{Brdar:2016ifs,Raffelt:2011nc,Arguelles:2016uwb}.

Neutrinos emitted from a Galactic SN are expected to reach Earth  about several hours ahead of the corresponding photons to which the SN interior is opaque.
From a statistical point of view, the next Galactic SN explosion will most likely happen in the Galactic plane where the stellar density peaks \cite{Mirizzi:2006xx}. This information, by far, does not allow us to pinpoint the location of the SN.
Moreover, SN detection via photon signals may fail due to the small field of view of satellites operating in the keV--MeV  energy range, or due to the absence of photon emission. The latter could happen if the SN core collapses into a black hole \cite{Beacom:2000qy,Burrows3}.
 The black hole produced in this process would yield a cut-off in the neutrino flux that corresponds to the moment of the black hole formation. In this scenario, the SN observation would heavily rely on the neutrino signal \cite{Burrows1,Wallace:2015xma}.

In view of the above aspects, a natural question emerges\textemdash``Can
SN location be robustly determined by using neutrinos?"
This approach, if possible, would have essential contributions to the Supernova Early Warning System \cite{Antonioli:2004zb}, which has been established to provide a timely indication of the next Galactic SN occurrence for  astronomical observations. Even in the presence of a strong optical signal, the potential complementary determination of SN location in neutrino detectors would be very valuable.

The idea of locating SN using neutrino signal was introduced in Ref. \cite{Beacom:1998fj} where two distinct options were presented. One option is to use the angular dependence of several neutrino interaction channels. Such an idea was also studied in \cite{Tomas:2003xn} and further explored in recent simulations performed by the Super-Kamiokande collaboration \cite{Abe:2016waf}. The alternative option proposed in \cite{Beacom:1998fj} was based on a multi-detector analysis via the triangulation method. The concept is that, based on the neutrino arrival time differences at three or more detectors, one can infer the direction of the incoming neutrino flux. Such a method was originally regarded to be relatively imprecise and not competitive with respect to the former option. However, with the significant development of neutrino detectors in the last decade, the triangulation method deserves reconsideration. Recently Ref. \cite{Muhlbeier:2013gwa} revisited this option and showed that the triangulation method can be robust
to locate Galactic SNe assuming the detectors reaching the sensitivity of 2 ms in the time differences. However, the triangulation approach still lacks more detailed studies regarding the method to determine time differences as well as an up-to-date global analysis.
The main purpose of this paper is to provide a full picture regarding the capabilities of all relevant current and near future neutrino detectors in order to determine SN location via the triangulation method. To this end, we numerically evaluate four different SN models provided by the Garching group \cite{garching} and perform  numerical fits in order to determine the uncertainties of the neutrino arrival times in all considered detectors. In our analysis, water Cherenkov, liquid scintillator as well as DM detectors are included.
Therefore, in this global picture, the flavor complementarity is also present as we consider both charged current and neutral current processes.

The remainder of this paper is organized as follows. In \cref{sec:SN-models} we introduce the SN neutrino flux
employed in our analysis and present the method to construct SN event rates in different experiments. Section \ref{sec:fitting} is dedicated to the numerical fits which are necessary for the precise determination of the uncertainties of the neutrino arrival times. These uncertainties are essential in order to obtain our main result, given in \cref{sec:results},
 where we show
 the robustness of the triangulation method by demonstrating how precise a Galactic SN can be located. Finally, we conclude in \cref{sec:summary}.


\section{Supernova Models and Event Rates}
\label{sec:SN-models}

In this section we introduce SN models to be used in our analysis and construct neutrino fluxes at Earth.
We discuss the most relevant detection channels for SN neutrinos and  describe the procedure to calculate the event rates in different experiments, which sets the stage for the subsequent analyses.

\subsection{SN Fluxes}
\label{subsec:fluxes}

\begin{figure}
  \centering
  \begin{tabular}{cc}
    \includegraphics[width=.48\textwidth]{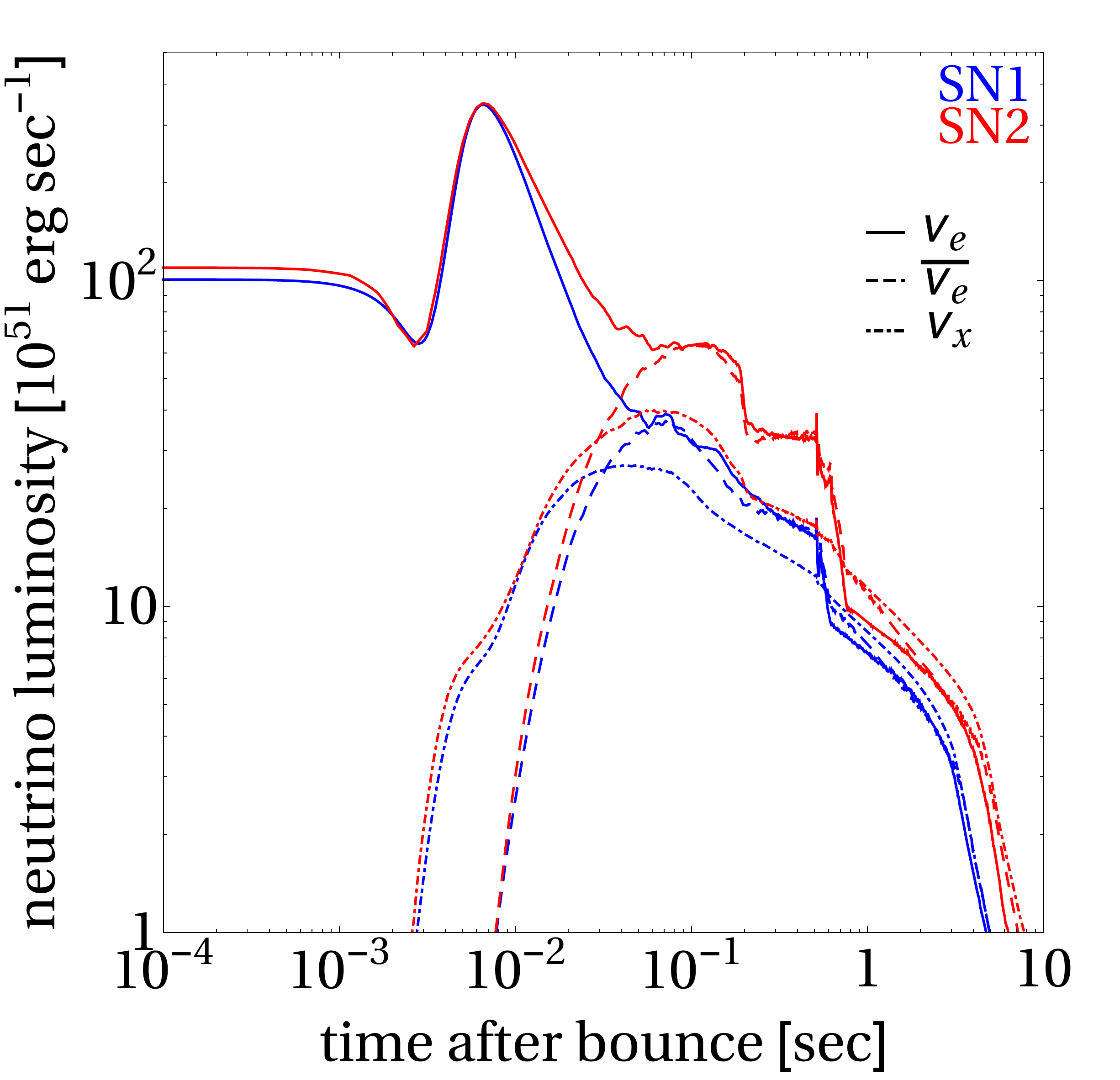} &
    \includegraphics[width=.48\textwidth]{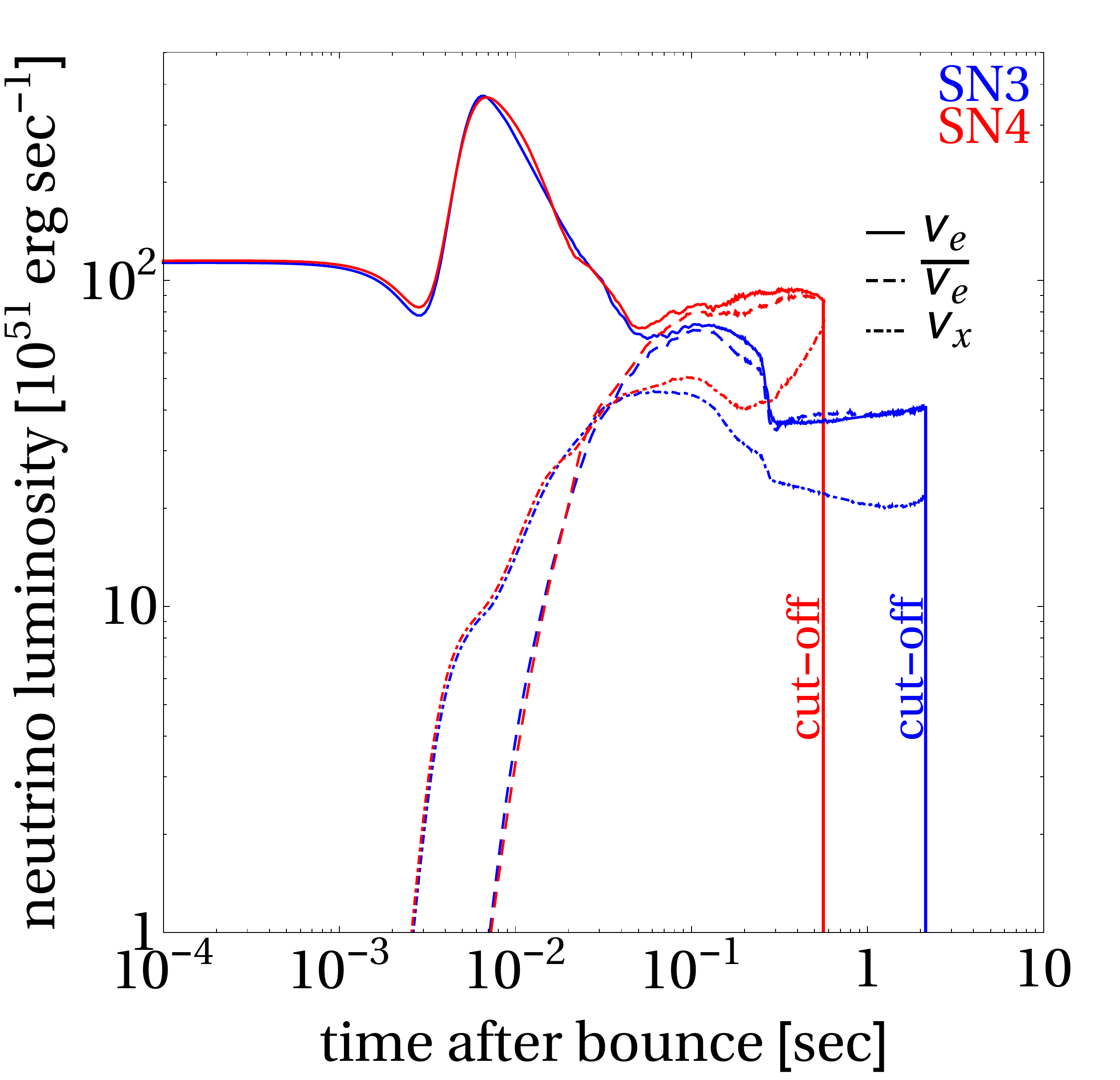} \\
     \\[0.2cm]
    \includegraphics[width=.48\textwidth]{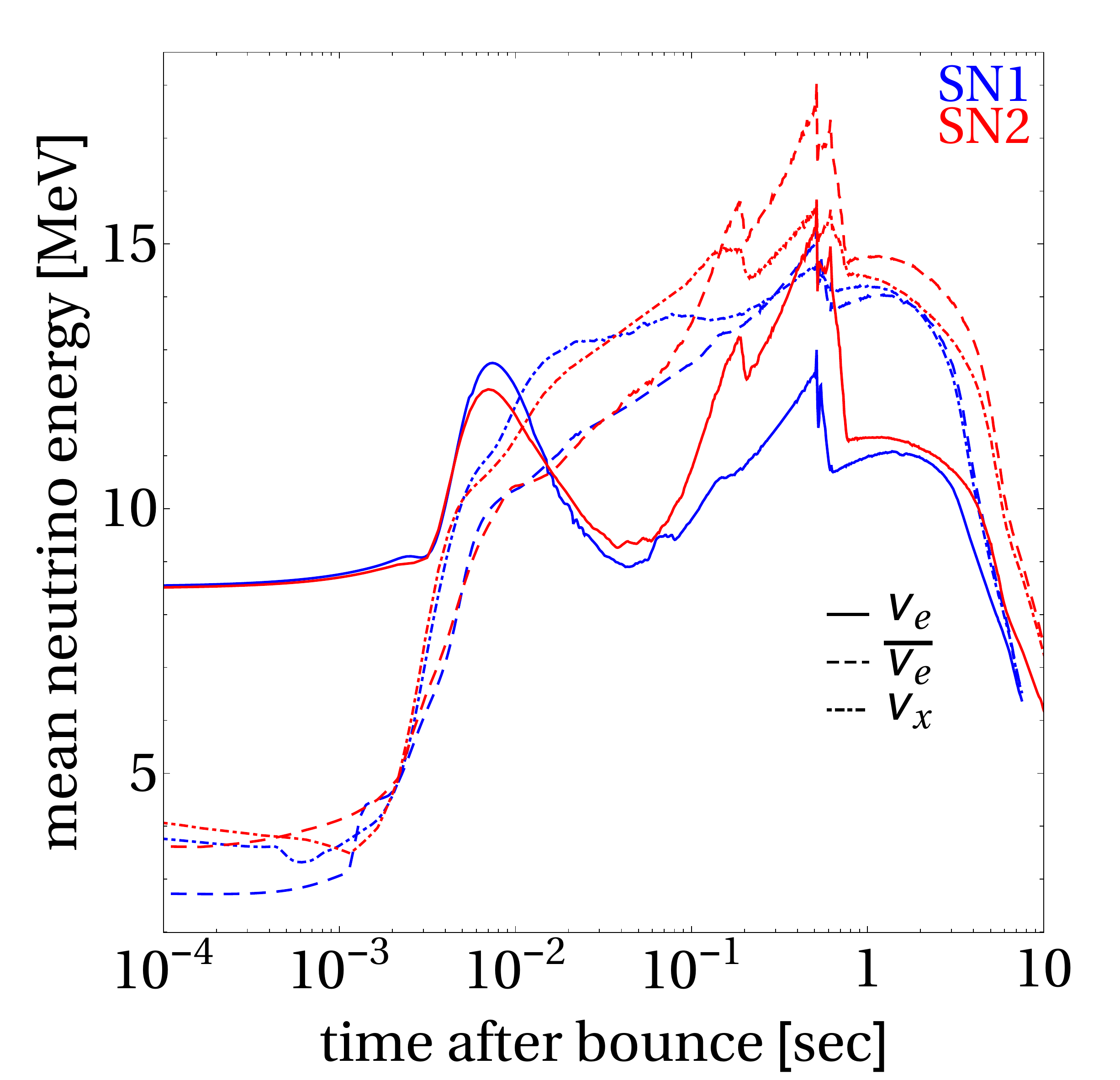} &
    \includegraphics[width=.48\textwidth]{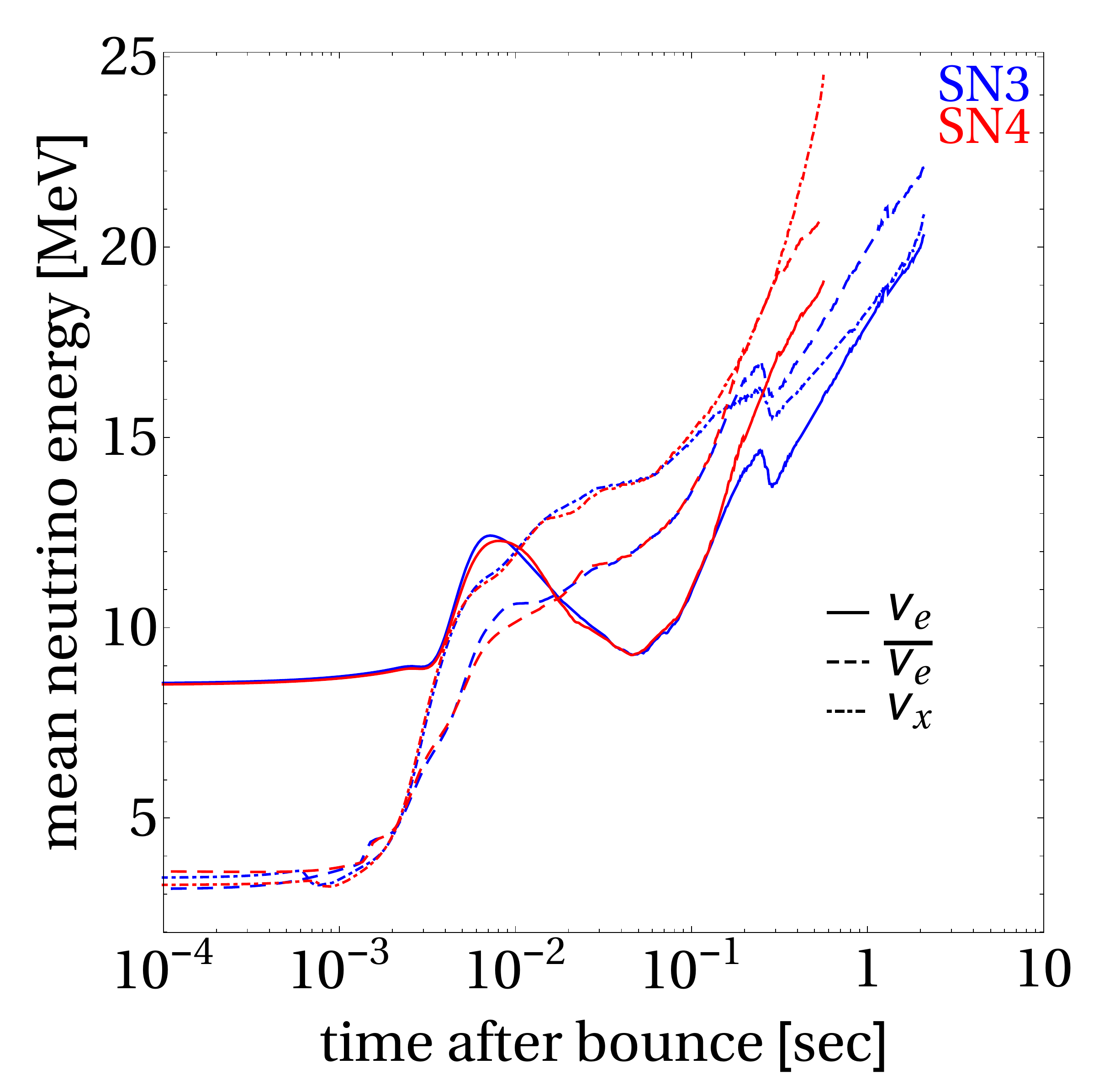} \\

  \end{tabular}
  \caption{Luminosities (upper panels) and mean energies (lower panels) for all neutrino flavors ($\nu_e$, $\bar{\nu}_e$ and $\nu_x$) produced in SNe. Figures on the left (right) are for SN1 and SN2 (SN3 and SN4) models. The prominent peak in the upper panels corresponds to the neutronization burst \cite{Kachelriess:2004ds}, followed by the accretion and cooling phase  when the vast majority of neutrinos is emitted. The total duration of these stages for the case of a neutron star formation after a core-collapse (SN1, SN2) is roughly 10 seconds \cite{Burrows2}. In a scenario where a black hole is formed (SN3, SN4), this time is significantly shorter and there is the rapid drop in the luminosity caused by a black hole formation.
  }
  \label{fig:fluxes+energies}
\end{figure}

We make use of machine-readable data underlying four different SN simulations performed by  the Garching group \cite{Hudepohl2013,Mirizzi:2015eza}. Two of them
are for 11 $M_\odot$ (SN1) and 27 $M_\odot$ (SN2) progenitor stars whose cores form neutron stars after the collapse, whereas the  remaining two correspond to $40 M_\odot$ progenitor stars yielding  black holes subsequent to the explosion (SN3 and SN4). Here, in brackets, for each of these scenarios we provide a corresponding abbreviation which will be used in the remainder of the paper. Note that the crucial difference between SN3 and SN4 is in the duration of neutrino signal, i.e. in the cut-off time coinciding with the black hole formation. This abrupt attenuation of neutrino fluxes is roughly at $2.1$  and $0.57$ seconds for SN3 and SN4, respectively, determined with respect to the SN core bounce.
In all these SN simulations, \emph{LS220} nuclear equation of state was employed. In
\cref{fig:fluxes+energies}, the neutrino fluxes and mean energies are shown as a function of time for all given scenarios at a distance of $500$ km from the center of the star. As there is virtually no difference in the spectral properties between neutrinos and antineutrinos of muon and tau flavors produced in SN (commonly denoted as $\nu_x$), we consider three different sets of fluxes and energies corresponding to electron neutrinos ($\nu_e$), electron antineutrinos ($\bar{\nu}_e$) and $\nu_x$.\\

The flux of (anti)neutrino flavor $\alpha$ at a distance $D$ from the center of the star is \cite{Keil:2002in}
\begin{align}
\phi_{\nu_\alpha}^0 (t,E)=\frac{\left[{1+\chi(t)}\right]^{1+\chi(t)}}{\langle E_{\nu_\alpha}(t)\rangle \,\Gamma\left[1+\chi(t)\right]} \frac{L_{\nu_\alpha}(t)}{4\pi D^2 \langle E_{\nu_\alpha}(t)\rangle } \left(\frac{E}{\langle E_{\nu_\alpha}(t)\rangle}\right)^{\chi(t)} \text{Exp}\left[-\frac{(1+\chi(t)) E}{\langle E_{\nu_\alpha}(t)\rangle}\right],
\label{eq:fluxes}
\end{align}
where $L_{\nu_\alpha}$ and $\langle E_{\nu_\alpha}\rangle$ are the luminosity and the mean energy of a given neutrino species $\alpha$, $\Gamma$ is the Gamma function and $\chi$ is the so called ``pinching" parameter which quantifies the deviation of the neutrino distribution function
with respect to the Maxwell-Boltzmann one. It is obtained via the following relation \cite{Keil:2002in}
\begin{align}
\frac{\langle E_{\nu_\alpha}^2(t) \rangle}{\langle E_{\nu_\alpha}(t)\rangle^2}=\frac{2+\chi(t)}{1+\chi(t)}.
\label{eq:pinching}
\end{align}
The expression in \cref{eq:fluxes} holds if the following condition for the density is satisfied
\begin{align}
\rho(D)\gtrsim 10^4 \, \text{g cm}^{-3}.
\end{align}

Around the radius where the density of the star is $10^3-10^4 \, \text{g cm}^{-3}$, the Mikheev-Smirnov-Wolfenstein (MSW) resonance \cite{Wolfenstein,Mikheev:1986gs} corresponding to the atmospheric neutrino mass squared difference ($|{\Delta m}_{13}^2|\approx 2.5\cdot 10^{-3} \,\text{eV}^2$) occurs. Another resonance which alters the flavor structure, associated with the solar mass squared difference (${\Delta m}_{12}^2\approx 7\cdot 10^{-5} \,\text{eV}^2$), is at densities of around $10\,\text{g cm}^{-3}$ \cite{Dighe:1999bi}. By using the established property of a practically fully adiabatic evolution of neutrinos through these resonances \cite{Dighe:1999bi}, one finds the following fluxes at a distance $d$ $(d\gg D)$ from the SN \cite{Dighe:1999bi,Lu:2016ipr}

\begin{align}
& \phi_{\nu_e}=\phi_{\nu_x}^0\left(\frac{D}{d}\right)^2, \nonumber \\ &
\phi_{\bar{\nu}_e}=\left(\phi_{\bar{\nu}_e}^0  \cos^2 \theta_{12} + \phi_{\nu_x}^0   \sin^2 \theta_{12} \right) \left(\frac{D}{d}\right)^2, \nonumber \\ &
\phi_{\nu_x}=\left(\frac{1}{4}\left(2+\cos^2 \theta_{12} \right) \phi_{\nu_x}^0 + \frac{1}{4} \phi_{\nu_e}^0+\frac{1}{4} \phi_{\bar{\nu}_e}^0 \sin^2 \theta_{12}\right) \left(\frac{D}{d}\right)^2,
\label{fluxes-normal}
\end{align}
for the normal ordering of neutrino masses and
\begin{align}
&\phi_{\nu_e}=\left(\phi_{\nu_e}^0  \sin^2 \theta_{12} + \phi_{\nu_x}^0   \cos^2 \theta_{12} \right) \left(\frac{D}{d}\right)^2, \nonumber \\ & \phi_{\bar{\nu}_e}=\phi_{\nu_x}^0\left(\frac{D}{d}\right)^2, \nonumber \\ &
\phi_{\nu_x}=\left(\frac{1}{4}\left(2+\sin^2 \theta_{12} \right) \phi_{\nu_x}^0 + \frac{1}{4} \phi_{\bar{\nu}_e}^0+\frac{1}{4} \phi_{\nu_e}^0 \cos^2 \theta_{12}\right) \left(\frac{D}{d}\right)^2,
\label{fluxes-inverted}
\end{align}
for the inverted one. In these equations, $\phi_{\nu_x}$  represents each individual (anti)neutrino flux of muon or tau flavor and $\theta_{12}\approx 33^\circ$ \cite{Esteban:2016qun}. What is explicitly taken into account in the derivation of
\cref{fluxes-normal,fluxes-inverted} is the absence of vacuum  neutrino oscillations outside of the star due to the wave packet decoherence effects. Throughout the paper we assume $d\approx 10$ kpc which is an approximate distance between the Galactic center and Earth. Finally, let us note that we ignore the possible flavor transitions coming from the collective effects \cite{Duan:2010bg} in the SN interior.

\subsection{Event Rates in Neutrino Detectors}
\label{subsec:ev-rates}

\begin{figure}
\centering

\includegraphics[width=0.9\textwidth]{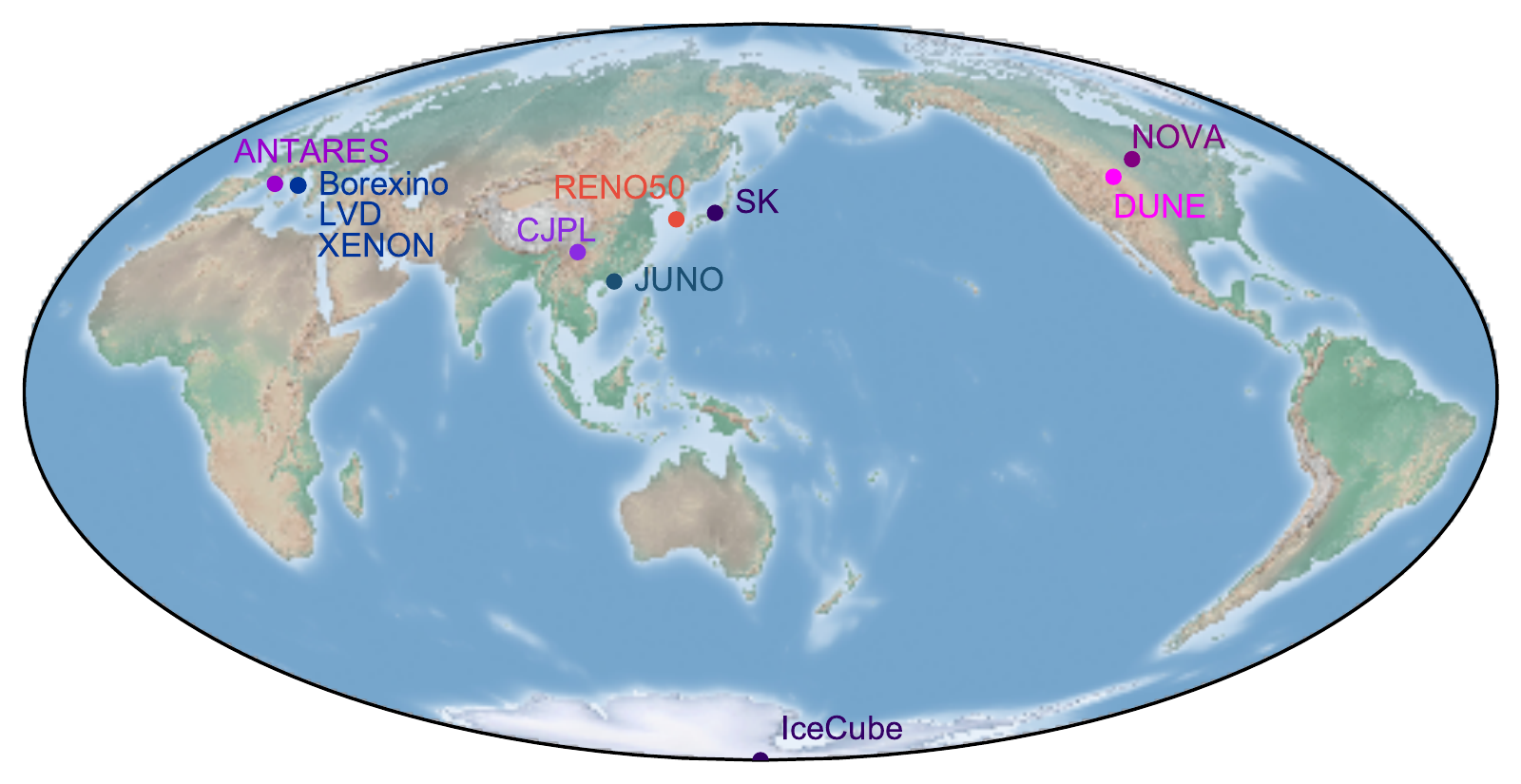}

\caption{The locations of the neutrino detectors considered in this paper shown on a world map.
}
\label{fig:detectors}
\end{figure}

With neutrino fluxes given in \cref{fluxes-normal,fluxes-inverted}, one can straightforwardly compute the expected number of SN (anti)neutrino events for a specific interaction channel.
For neutrinos of flavor $\alpha$, the number of events in the $i$-th time bin is

\begin{equation}
N_{\alpha}^{i}(A,\thinspace t_{0})=A\int dE \int_{t_{i}}^{t_{i}+\Delta t}\,dt\,\sigma(E)\,\phi_{\nu_{\alpha}}(\thinspace t-t_{0}, E),
\label{eq:event-number}
\end{equation}
where $A$ is the total number of target particles, $\sigma$
is the cross section for a given detection process, $\phi_{\nu_{\alpha}}$ represents (anti)neutrino
flux of flavor $\alpha$  and $E$ is the neutrino energy that is integrated over.
We divide the total time interval into a vast number of bins where the $i$-th bin lies in the interval $[t_{i},\thinspace t_{i}+\Delta t]$.
In \cref{eq:event-number}, we have also inserted a time shift, $t_0$, which is defined as the time delay with respect to a certain reference point that can be set arbitrarily. For example, one can define $t_0$ as the time delay of neutrino arrival to the detector with respect to the time when the neutrino flux crosses the center of Earth.
Since for our analysis only the time differences in SN neutrino detection in worldwide distributed detectors (see \cref{fig:detectors}) are relevant, a global $t_0$ shift has no observable effect.\\

The most efficient channel for the detection of SN neutrinos is the inverse beta decay (IBD)\cite{Vogel:1999zy}
\begin{equation}
\overline{\nu}_{e}+p\rightarrow e^{+}+n,
\label{eq:IBD}
\end{equation}
which has a threshold at much lower energy (1.8 MeV) with respect to the typical SN neutrino energies. The successful background suppression stems from a correlation in timing between prompt $\gamma$ from $e^+ e^-$ pair annihilation and a delayed $\gamma$ from neutron capture.

IBD is the dominant process
to detect
SN neutrinos in  water Cherenkov detectors (current Super-Kamionande \cite{Abe:2016waf,Laha:2013hva} and future Hyper-Kamiokande \cite{Yokoyama:2017mnt}), as well as in
liquid scintillators. In the latter, neutrino scattering on protons may be  relevant, depending on the proton recoil energy threshold of a given experiment. For instance, when compared to the expected IBD event rate, neutrino--proton scattering in JUNO \cite{An:2015jdp,Lu:2016ipr,Laha:2014yua} yields 33 \%, in  Borexino \cite{Lujan-Peschard:2014lta}  23\%, and in Kamland \cite{Lujan-Peschard:2014lta} 11\% of the total IBD event numbers, whereas in NOvA \cite{Vasel:2017egd} it is negligibly small.
Despite being clearly subdominant with respect to IBD, this channel has the advantage of being flavor independent since it is a neutral current process. This may be very relevant for the studies dedicated to SN neutrino flavor compositions. As for our purpose, we have checked that the estimates for the uncertainties of the neutrino arrival times, discussed in the following section, are basically insensitive to this channel.
Therefore, for liquid scintillator detectors, we take IBD as the only relevant channel. Let us note here that IBD is the main detection channel also for the neutrino telescopes such as  IceCube \cite{Aartsen:2016nxy} and Antares \cite{Ageron:2008hv} as well as future KM3NeT \cite{Adrian-Martinez:2016fdl}.\\

The detectors which would virtually only probe SN electron neutrinos ($\nu_e$) are those containing  liquid argon (LAr). The relevant process is the charged-current quasi-elastic process (CCQE) on ${\rm ^{40}Ar}$:
\begin{equation}
\nu_{e}+{\rm ^{40}Ar}\rightarrow e^{-}+{\rm ^{40}K^{*}}.
\label{eq:Lar}
\end{equation}
Currently, there is MicroBooNE \cite{Soderberg:2009rz} which is not designed to capture SN neutrinos efficiently and  would detect only $\sim 20$ events from $10$ kpc distant SN. However, a future project---DUNE \cite{Acciarri:2016crz} will consist of 40 tonnes of LAr and will be able to robustly probe SN \cite{Nikrant:2017nya}. The expected number of $\nu_e$ events from SN in the Galactic center is $\sim 3000$. We include DUNE far detector, but omit MicroBooNE from our analysis due to its insufficient sensitivity to SN.\\

Current DM detectors have reached both technology and size to guarantee detection of  neutrinos from Galactic SN \cite{Lang:2016zhv,Chakraborty:2013zua} via coherent neutrino--nucleus scattering.

We compute the number of neutrino events at DM experiments by employing
\cref{eq:event-number}, where $A$ is now the number of Xenon atoms present in the detector and the cross section $\sigma_\text{\text{coh}}$ can be written as

\begin{align}
\sigma_\text{\text{coh}}(E)= \int_{E_\text{th}} dE_{\text{recoil}}  \frac{d\sigma}{dE_{\text{recoil}}},
\label{eq:DM-xsec}
\end{align}
  where the integral in the recoil energy $E_{\text{recoil}}$ runs from its threshold value $E_\text{th}$. The differential cross section is \cite{Lang:2016zhv}

\begin{align}
\frac{d\sigma}{dE_{\text{recoil}}}=\frac{G_F^2 m_{\text{Xe}}}{4\pi} \left[N_n-(1-4\sin^2 \theta_W) N_p\right]^2 \left(1-\frac{m_{\text{Xe}} E_\text{recoil}}{2 E^2}\right) F^2(E_\text{recoil}),
\end{align}
where $G_F$ is the Fermi constant, $\theta_W$ is the weak mixing angle, $N_n$, $N_p$ and $m_\text{Xe}$ are neutron number, proton number and the mass of a xenon atom, respectively, and $F(E_\text{recoil})$ is the nuclear form factor for which we adopt the expression from Refs. \cite{Vietze:2014vsa,Lang:2016zhv}. Note that
the most general formula for the event number would  include the detector response in the form of two scintillation signals \cite{Lang:2016zhv}. For SN1 and SN2 models, our approximation in \cref{eq:event-number} accurately reproduces results  given in Ref. \cite{Lang:2016zhv}, for $E_\text{th}=1$ keV. In order to provide
simulated
event numbers we again use Poissonian statistics.
At XENON1T \cite{Aprile:2017aty} experiment $\mathcal{O}(10)$  events are obtained (see \cref{tab:detect}). We consider
XENON1T in our triangulation analysis. Due to the non-competitive event numbers with respect to the larger neutrino detectors, it yields a rather marginal impact. However, its future 40-tonne successor DARWIN \cite{Aalbers:2016jon} has better prospects. Analogously to the previously discussed neutrino-proton scattering in liquid scintillators, in DM detectors all flavors have identical cross section for coherent scattering on nuclei. Hence, the effective difference in the number of events across different neutrino species stems exclusively from the flux.

\begin{figure}
\centering

\includegraphics[width=0.45\textwidth]{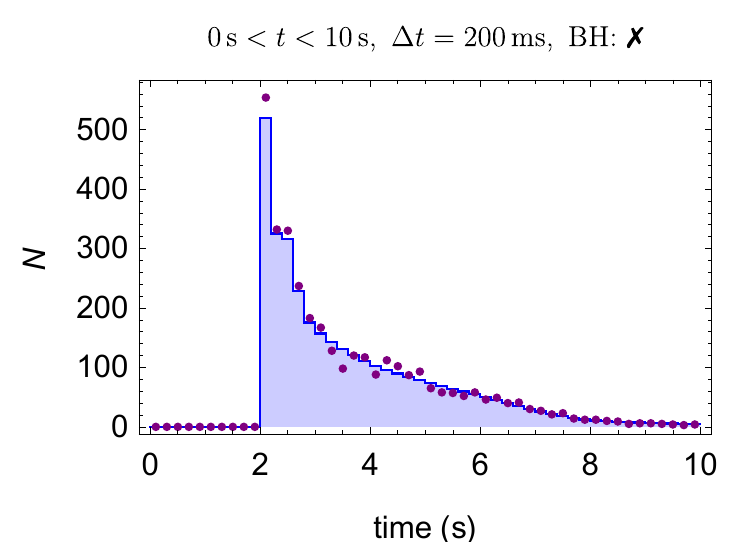}\ \includegraphics[width=0.45\textwidth]{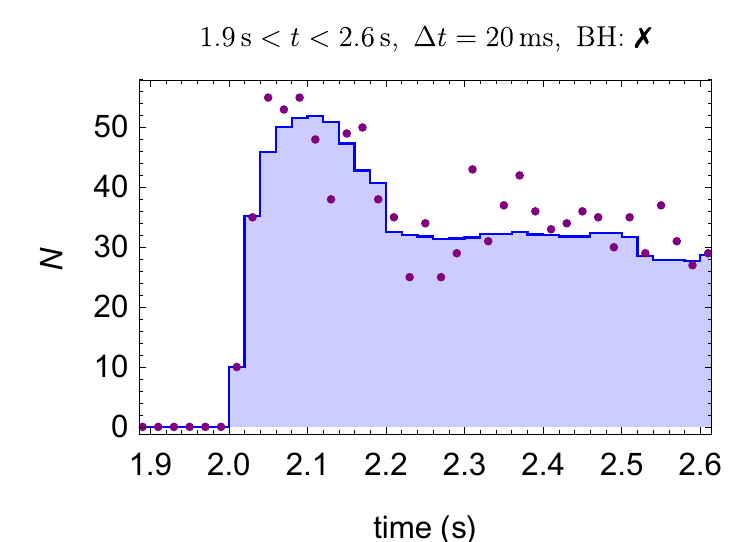}

\includegraphics[width=0.45\textwidth]{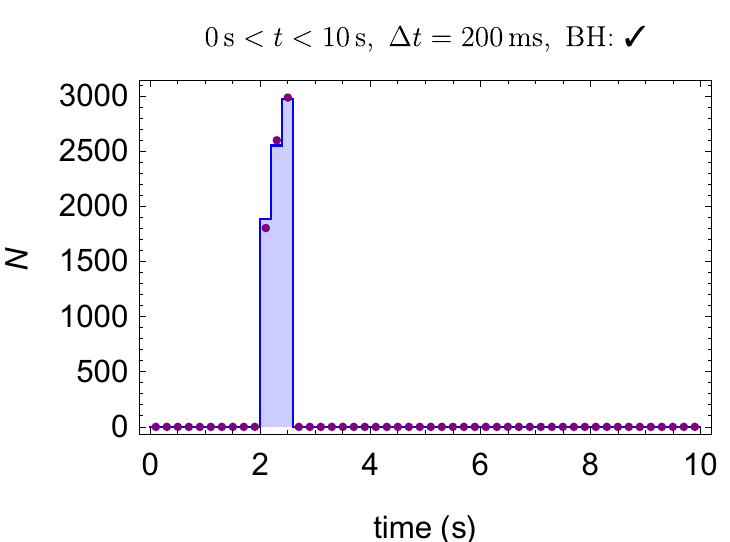}\ \includegraphics[width=0.45\textwidth]{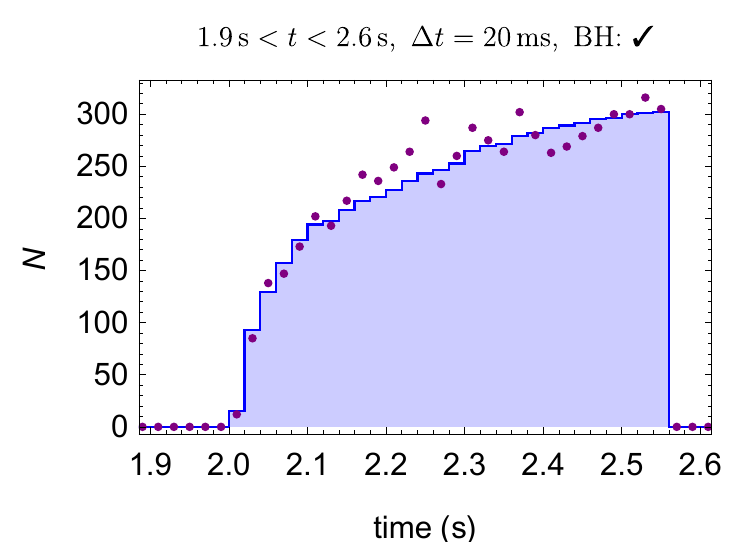}\

\caption{\label{fig:events1}SN $\bar{\nu}_e$  events at the Super-Kamiokande detector. The neutrinos are emitted from a 10 kpc distant SN. In the upper (lower) panels  SN2 (SN4) model is considered. In addition, the labels ``BH:\cmark" and ``BH:\xmark"
refer to whether a black hole is formed (SN4) or not (SN2). Parameter $t_{0}$ is fixed to $2.0$ sec. Expected event distributions are computed
in two time windows: $0\thinspace{\rm s<}t<10\thinspace{\rm s}$ (left); $1.9\thinspace{\rm s<}t<2.6\thinspace{\rm s}$ (right).
The time windows and $\Delta t$ used in the figure are only for illustration. For actual data fitting, we use much wider time windows and much smaller $\Delta t$ as explained in Sec.~\ref{sec:fitting}.
 The blue histograms indicate the event rates computed by using \cref{eq:event-number}, whereas the purple points are generated according
to \cref{eq:Poisson} in order to realistically simulate the observed events.
}
\end{figure}

\begin{figure}
\centering

\includegraphics[width=0.45\textwidth]{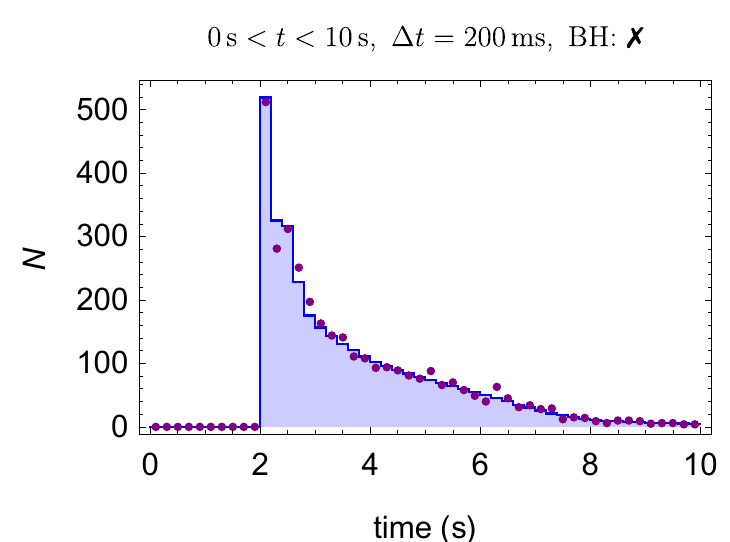}\ \includegraphics[width=0.45\textwidth]{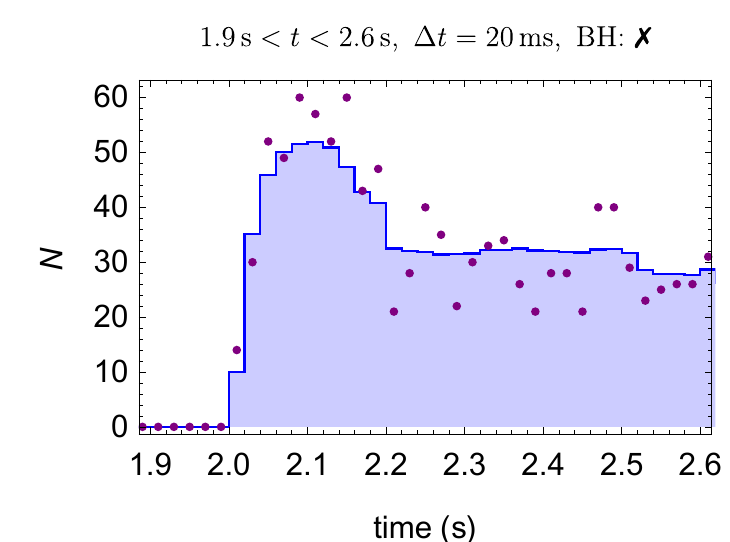}

\includegraphics[width=0.45\textwidth]{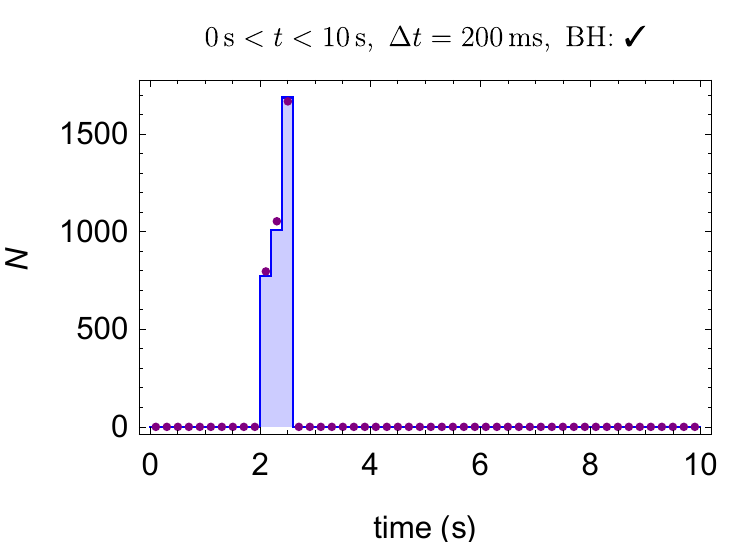}\ \includegraphics[width=0.45\textwidth]{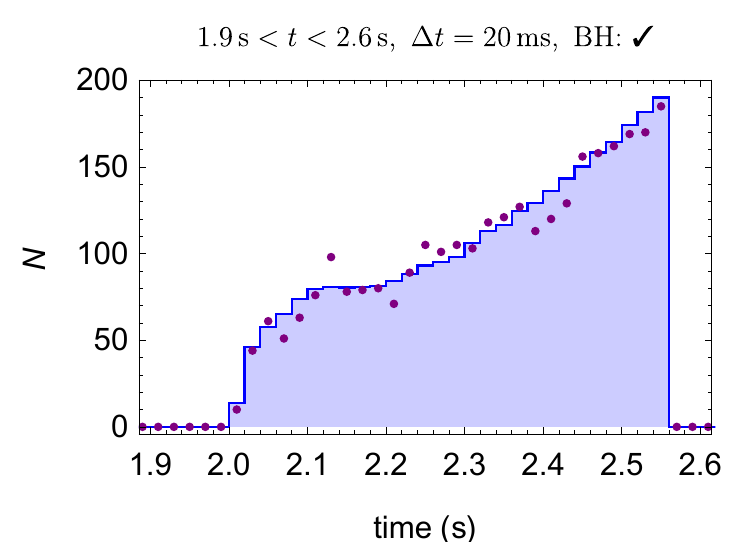}\

\caption{\label{fig:eventsDUNE}
SN $\nu_e$ event rates at DUNE considered for the scenarios identical to those in \cref{fig:events1}.
}
\end{figure}

In order to illustrate the calculation, let us first take Super-Kamiokande detector as an example. As already discussed, the main  SN detection channel in  Super-Kamionande is IBD  and we focus on that channel only. The fiducial mass of the detector is 32 kt of ${\rm H}_{2}{\rm O}$ containing
\begin{equation}
A=32\thinspace{\rm kt}\frac{2N_{A}}{18.015\thinspace{\rm g/mol}}=2.14\times10^{33}
\label{eq:target-number}
\end{equation}
hydrogen nuclei (i.e. free protons). In the expression, $N_{A}=6.02\times10^{23}$ $\text{mol}^{-1}$
is the Avogadro constant. We set $t_{0}=2.0$ sec at the point when
SN neutrinos arrive at the detector and compute the event distributions
 in the following two time windows
\begin{itemize}
 \item $0\thinspace{\rm s<}t<10\thinspace{\rm s}$,\,
$\Delta t=200\thinspace{\rm ms}$,
\item $1.9\thinspace{\rm s<}t<2.6\thinspace{\rm s}$,\,
$\Delta t=20\thinspace{\rm ms}$.
\end{itemize}
One should keep in mind that these values are only for illustration while in the actual evaluation of the time uncertainties (in Sec. III) we need to use wider time windows and much smaller $\Delta t$, which are not suitable for proper visualization in Figs.~\ref{fig:events1} and \ref{fig:eventsDUNE}.

 We adopt the IBD cross section from Ref. \cite{Strumia:2003zx} and the $\overline{\nu}_{e}$
flux at Earth for the assumed case of normal neutrino mass ordering.
The results for the case of a neutron star (SN2) and a black hole (SN4) formation are shown in \cref{fig:events1}
where the blue histograms indicate the event rates computed by using \cref{eq:event-number}. The other two models, SN1 and SN3, do not provide a qualitative distinction in the obtained event rates as well as in the spectral shape and are therefore not shown.

In order to simulate the observed integer number of events per bin, we also show purple points in \cref{fig:events1}, generated by the Poisson distribution
\begin{equation}
P=\frac{\mu^{n}}{n!}e^{-\mu},
\label{eq:Poisson}
\end{equation}
which gives the probability of detecting $n$ events in a bin with an expectation value $\mu=N_{\alpha}^i$ computed via \cref{eq:event-number}.\\

The procedure for DUNE follows in a similar fashion, with the difference in the cross section. For the process given in \cref{eq:Lar} we adopt the cross section values from \cite{Gil-Botella:2016sfi}. The SN neutrino events for DUNE are shown in \cref{fig:eventsDUNE} in an analogous way as for the Super-Kamiokande (see
\cref{fig:events1}).\\

\section{Uncertainties of Supernova neutrino arrival time}
\label{sec:fitting}

\begin{table*}
\begin{ruledtabular}
\begin{tabular}{ccccccc}
Experiments & major process & target & $N_{{\rm total}}$ & $\delta t$  & $N_{{\rm total}}$ (BH) & $\delta t$ (BH)\tabularnewline
\hline
Super-Kamiokande \cite{Abe:2016waf} & $\overline{\nu}_{e}+p\rightarrow e^{+}+n$ & 32 kt ${\rm H}_{2}{\rm O}$ & 7625 & 0.9 ms & 6666 & 0.14 ms\tabularnewline
JUNO \cite{An:2015jdp} & $\overline{\nu}_{e}+p\rightarrow e^{+}+n$ & 20kt ${\rm C}_{n}{\rm H}_{m}$ & 4766 & 1.2 ms & 4166 & 0.19 ms\tabularnewline
RENO50 \cite{Kim:2014rfa} & $\overline{\nu}_{e}+p\rightarrow e^{+}+n$ & 18kt ${\rm C}_{n}{\rm H}_{m}$ & 4289 & 1.3 ms & 3749 & 0.21 ms\tabularnewline
DUNE \cite{Acciarri:2015uup} & $\nu_{e}+{\rm ^{40}Ar}\rightarrow e^{-}+{\rm ^{40}K^{*}}$ & 40 kt LAr & 3297 & 1.5 ms & 3084 & 0.18 ms\tabularnewline
NO$\nu$A \cite{Patterson:2012zs} & $\overline{\nu}_{e}+p\rightarrow e^{+}+n$ & 15 kt ${\rm C}_{n}{\rm H}_{m}$ & 3574 & 1.4 ms & 3125 & 0.24 ms\tabularnewline
CJPL \cite{Jinping2016} & $\overline{\nu}_{e}+p\rightarrow e^{+}+n$ & 3kt ${\rm H}_{2}{\rm O}$ & 715 & 3.8 ms & 625 & 0.97 ms\tabularnewline
IceCube \cite{Halzen:2010yj} & noise excess & ${\rm H}_{2}{\rm O}$ & $\mathcal{O}(10^6)$\cite{ICdeltat} & 1ms  & $\mathcal{O}(10^6)$\cite{ICdeltat} & 0.16 ms \tabularnewline
ANTARES \cite{Sokalski:2005sf} & noise excess & ${\rm H}_{2}{\rm O}$ & $\mathcal{O}(10^3)$\cite{Kulikovskiy:2017pkv} & 100ms & $\mathcal{O}(10^3)$\cite{Kulikovskiy:2017pkv} & 32 ms\tabularnewline
Borexino \cite{Alimonti:2000xc} & $\overline{\nu}_{e}+p\rightarrow e^{+}+n$ & 0.3 kt ${\rm C}_{n}{\rm H}_{m}$ & 71.5 & 16 ms & 62.5 & 5.5 ms\tabularnewline
LVD \cite{Aglietta:2002gj}& $\overline{\nu}_{e}+p\rightarrow e^{+}+n$ & 1 kt ${\rm C}_{n}{\rm H}_{m}$ & 238 & 7.5 ms & 208 & 2.4 ms\tabularnewline
XENON1T \cite{Aprile:2017iyp} & coherent scattering & 2t ${\rm X_{e}}$ & 31 & 27 ms & 29 & 10 ms\tabularnewline
DARWIN \cite{Aalbers:2016jon} & coherent scattering & 40t ${\rm X_{e}}$ & 622 & 1.3 ms & 588 & 0.7 ms\tabularnewline
\end{tabular}\end{ruledtabular}
\caption{Global picture for SN neutrino event rates and arrival time
uncertainties ($\delta t$). In the second and third columns, the dominant detection channel as well as the target are indicated for each of the experiments shown in the first column.
The next (last) two columns show total event numbers $N_{\rm{total}}$ and $\delta t$ values for
the core collapsing into a neutron star (black hole). The values are computed based on Eq.~(\ref{eq:event-number}) and Eq.~(\ref{eq:sn-7}), except for IceCube and ANTARES for which we refer the reader to the discussion in the text.}

\label{tab:detect}
\end{table*}

\begin{table*}

\begin{ruledtabular}
\begin{tabular}{ccccc}
total events & SN1 (11 $M_{\odot}$) & SN2 (27 $M_{\odot}$) & SN3 (BH) & SN4 (BH)\tabularnewline
\hline
$10^{5}$ & 0.2 ms & 0.2 ms & 0.06 ms & 0.02 ms\tabularnewline
$10^{4}$ & 0.8 ms & 0.8 ms & 0.3 ms & 0.1 ms\tabularnewline
$10^{3}$ & 2.9 ms & 3.1 ms & 1.9 ms & 0.6 ms\tabularnewline
$10^{2}$ & 11 ms & 13 ms & 7.3 ms & 4 ms\tabularnewline
\end{tabular}\end{ruledtabular}
\caption{$\delta t$  for IBD detectors of different sizes. This table can also be used to approximately evaluate $\delta t$ for LAr detectors and DM detectors provided the total even numbers, but it is not applicable to IceCube and ANTARES which detect SN neutrinos via noise excess.}
\label{tab:uncertainty}
\end{table*}

In this section we introduce the chi-square goodness of fit method in order to determine the time delay $t_0$ for all relevant current and near-future detectors, and present the corresponding statistical uncertainties ($\delta t$).
The full list of experiments is given in \cref{tab:detect}.

Here we would like to further clarify the meaning of $t_0$ and $\delta t$. One should note that we do \emph{not} tag any special time point such as the peak of the flux, the arrival time of the first neutrino, or the onset of SN explosion, etc. We only define $t_0$ as the time delay of the arrival of a neutrino (any single neutrino amidst the neutrino flux) at the detector with respect to the time when the neutrino has passed through a reference point that can be set arbitrarily at a certain place, for instance the center of Earth. Even in the case when the neutrino flux varies very slowly in time, its time shift with respect to the reference point is still unambiguously  defined as long as there is no spectral distortion in the last few fractions of a second. The statistical uncertainty of measuring $t_0$ is defined as $\delta t$, which then becomes independent of the choice of the reference point.

As pointed out in Ref. \cite{Beacom:1998fj}, the minimal statistical uncertainties
on the neutrino arrival time can be estimated by employing the Cramer-Rao theorem \cite{Hogg}
\begin{equation}
\frac{1}{(\delta t)^{2}}=N\int dt\,\frac{[f'(t)]^{2}}{f(t)},
\label{eq:sn-8}
\end{equation}
where $f(t)$ is the normalized event rate distribution function, $f'(t)$ is its time derivative and $N$ represents the total number of detected events. This method can not be applied
to cases when $f(t)$  experiences an instant growth or a drop, such as in
the black hole formation scenario. Moreover, the impact of the background
can not be included in such an approach. Therefore, we adopt a generally applicable
$\chi^{2}$ fit in order to compute  $\delta t$.
We have checked that, for the case of a neutron star formation after the core-collapse (models SN1 and SN2), our results
agree with the values obtained from \cref{eq:sn-8} in the limit of low background
and high statistics. In what follows, we discuss the details of the fit.\\

For a large number of SN events, one can perform binning of the data in time frames (see for instance \cref{fig:events1}). The
$\chi^{2}$ value can be computed via
\begin{equation}
\chi^{2}(t_{0})=2\sum_{i=1}^{i_{{\rm max}}}\left(\mu_{i}-n_{i}+n_{i}\ln\frac{n_{i}}{\mu_{i}}\right),
\label{eq:sn-7}
\end{equation}
where $n_{i}$ ($\mu_{i}$) are the observed (expected)
event numbers computed by using \cref{eq:event-number,eq:Poisson} and $i_{{\rm max}}$ is the number of time bins considered in the $\chi^2$-fit.
Note that the parameter $t_{0}$ implicitly enters into $\chi^{2}$. This is because when $t_{0}$ shifts, $\mu_{i}$ varies correspondingly
according to \cref{eq:event-number}.

After data binning, depending on the bin width $\Delta t$, there may be a certain loss of the timing information. Therefore, for the sake of the precision,
$\Delta t$ must be much smaller than the statistical uncertainty $\delta t$, i.e.
\begin{equation}
\Delta t\ll\delta t.
\label{eq:sn-6}
\end{equation}

For a very small $\Delta t$, the event numbers in each bin (both $\mu_{i}$ and
$n_{i}$) are correspondingly small. We
determined that generally in order to reach the requirement given in \cref{eq:sn-6}, $\mu_{i}$ are in most cases smaller than $\mathcal{O}(1)$ and $n_{i}$, being integer numbers,
are mostly zero or one, though lager values may appear. Note that \cref{eq:sn-7} is still valid for such small values of $\mu_{i}$ and
$n_{i}$ since it is derived based on the Poisson distribution which
is particularly applicable for low statistics.
 Actually, in the $\Delta t\rightarrow0$ limit,
one recovers the case of fitting to the original unbinned data, i.e.
a series of events with a sequence in time. Despite $\Delta t\rightarrow0$ limit being   very difficult to probe, due to the extremely large number
of bins, we have found the width of $\Delta t$ for which
$\chi^{2}$ value is stable, in a sense that further reduction in $\Delta t$
does not lead to the observable difference in $\chi^2$. Needless to say,
for such $\Delta t$  \cref{eq:sn-6} is satisfied.

In order to perform a more realistic study, we include a background $\mu_{\rm{bkg}}$ in the $\chi^2$-fit by adding $\mu_{\rm{bkg}}$ to \cref{eq:event-number}.
For Super-Kamiokande, we take $\mu_{{\rm bkg}}=0.01\Delta t/{\rm s}$, i.e. 0.01 events per second, adopted from Ref. \cite{Abe:2016waf}. For other experiments, the background is simply rescaled
according to the fiducial mass of a detector.

 It should be noted
that one of the experiments in our analysis, NO$\nu$A \cite{Patterson:2012zs}, is not underground and suffers from a larger cosmic background. Therefore, the background for NO$\nu$A  in our analysis is underestimated.
A dedicated analysis to NO$\nu$A is beyond the scope of this paper and we refer to Ref. \cite{Vasel:2017egd} for more details.

When using \cref{eq:sn-7}, in principle, one should include the total
 SN neutrino signal in the time interval in which the
 fitting is performed.
Namely, the first bin should correspond to the time window before the burst and the last
bin should be set to the time after the end of neutrino signal.
In the neutron star formation scenario, the
 neutrino emission typically leaves a long tail. So technically $i_{\max}$ corresponds
to a cut at a certain value.
We have checked that the $\chi^{2}$ fit is practically
insensitive to the cut. For example, if $i_{\max}$ is set at earlier times
when only 90\% events are collected, we find the  negligible shift in $\chi^{2}$.
 This is because the main impact on the determination of $t_{0}$ comes
from  the drastic variation of the number of events between the neighboring bins. Therefore, for the neutron star formation case,
$t_{0}$ is almost exclusively determined by the first few percent of the events.
If the core collapses to a black hole, the flux has a very sharp cutoff as we already shown in \cref{fig:fluxes+energies}.
According to Ref. \cite{Beacom:2000qy}, the duration
of the cutoff could be around 0.5 ms. To study how this uncertainty affects our results, we added a $0.5$ ms linear tail in SN3 and SN4 models and found a slight increase in $\delta t$ by  $\sim0.04$ ms. Therefore, the tail does not pose a significant effect on $\delta t$. To be conservative, we include such tail in all calculations when the black hole is formed.
Let us note that for the black hole case $\delta t$ is almost exclusively determined
by the last couple of percent of events. In other words, the cut-off yields a more significant statistical effect in comparison to the rise
of flux at the signal onset.\\

We obtain $\delta t$ from the $\chi^{2}$ function by calculating the $1\sigma$ upper
and lower
deviations from its minimal value.
When these two ranges are unequal,  we conservatively associate the larger one with $\delta t$.
For the black hole case, we find that the $\chi^{2}$ function
is highly non-Gaussian. Hence, we adopt a different definition of $\delta t$
which is closer to the notion of Gaussian uncertainties. We first compute
the uncertainty at 3$\sigma$ ($\Delta\chi^{2}=9$) and divide it
by 3 in order to reach $\delta t$.
In Gaussian cases, this is identical to the $1\sigma$ bound. For non Gaussian cases, by applying this procedure, we  conservatively avoid an overoptimistic estimation of  $\delta t$.

As already mentioned, the event numbers $n_{i}$  are
 randomly generated with the Poisson distribution and the $\chi^{2}$ fit
is obtained by using \cref{eq:sn-7}. In this
approach, $\delta t$ has random fluctuations. To avoid them,
one can take the average value of $\delta t$ after many repetitive simulations. We find
that this value is approximately the same as the one computed by replacing
$n_{i}$ with the corresponding expected number of events $\mu_{i}=\langle n_i\rangle$.
Therefore, the
$\delta t$ values presented in this section are all computed by using the
average values of $n_{i}$, instead of averaging over many simulations.

The calculated values of $\delta{t}$
are summarized in \cref{tab:detect} for all considered experiments.
We select  SN2 and SN4
as representatives for the core collapsing into neutron stars and
black holes respectively, as for SN1 and SN3 models we did not find qualitative
differences. We assume normal ordering of neutrino masses and a 10 kpc distant SN.
 Larger or shorter distances
would certainly lead to different results. Since $\delta t$
mainly depends on the event numbers rather than the interaction channels for the detection,
we also present $\delta t$ for several selected total
event numbers (assuming IBD process)  in
\cref{tab:uncertainty}. For SN distances different than the one assumed,
and, in addition, for detectors not listed in \cref{tab:detect}, one can roughly
estimate $\delta t$ by using \cref{tab:uncertainty}.

In this work, we do not compute event rates, and consequently $\delta t$, for IceCube and ANTARES which detect SN neutrinos via noise excess. However, robust analyses, in particular for IceCube have been performed by the experiment collaborations. We infer from  \cite{Kopke:2017req,ICdeltat,ICdeltat2,Kulikovskiy:2017pkv} that $\delta t\sim 1$ ms for IceCube and $\delta t \sim 100$ ms for ANTARES,
assuming the core collapsing into a neutron star.  For the black
hole case, we multiply these values by $0.16$ and $0.32$, respectively  which
are the ratios between $\delta t$ values for  SN4 and SN2 models calculated for other experiments. 

Finally, let us comment that the Hyper-Kamiokande \cite{Abe:2011ts} experiment, a future successor of Super-Kamiokande will have a world leading $\delta t$ precision which may be
read off from \cref{tab:uncertainty} for the total event number of around $\mathcal{O}(10^5)$.
Despite such high precision, it has only little advantage in our analysis with respect to Super-Kamiokande  since in the triangulation method (see \cref{sec:results}) only the larger $\delta t$ between each pair of detectors is relevant.
\section{Triangulation Results}
\label{sec:results}

The time difference of SN neutrino arrival at two detectors located at $\vec r_i$ and $\vec r_j$ is \cite{Beacom:1998fj,Muhlbeier:2013gwa}

\begin{align}
t_{ij}=\frac{(\vec r_i-\vec r_j)\cdot \vec{n}}{c},
\label{eq:time-dif}
\end{align}
where $\vec n$ denotes the unit vector pointing to the direction of incoming neutrinos and, for definiteness, we choose the coordinate system with the origin in the center of Earth. In this equation, we ignore subleading effects caused by the nonvanishing masses of neutrinos, namely  the wave packet separation between the propagating mass eigenstates.

Due to Earth's rotation, the coordinates of the detectors ($\vec r_k$) are time dependent. Another time dependent correction stems from the revolution around the Sun. For simplicity, following Ref. \cite{Muhlbeier:2013gwa}, we assume SN observation on the vernal equinox at noon. Then, the angular coordinates of the detectors are simply the longitude $\phi_\text{Lon}$ and the latitude $\phi_\text{Lat}$ in the world geodetic system (WGS) \cite{wgs84} and we have

\begin{align}
&{r_k}^x= R_\oplus \cos \phi_\text{Lon} \cos \phi_\text{Lat}, \nonumber \\ &
{r_k}^y= R_\oplus \sin \phi_\text{Lon} \cos \phi_\text{Lat}, \nonumber \\ &
{r_k}^z  =  R_\oplus  \sin \phi_\text{Lat},
\label{coordinates}
\end{align}
 in the Cartesian coordinate system where $\vec r_k=({r_k}^x,{r_k}^y,{r_k}^z)$ and $R_\oplus$ is the Earth radius. For consistency, we employ the Earth-centered equatorial coordinate system with right-ascension $\alpha$ and declination $\delta$ to measure SN location in the sky. The unit vector $\vec n$ in this coordinate system is $\vec n=(-\cos \alpha \cos\delta, -\sin \alpha \cos \delta, -\sin \delta)$, where the minus sign indicates the direction of the incoming neutrinos. The conversion between the more commonly used galactic coordinate system and the equatorial one is straightforward (see Ref. \cite{Mirizzi:2006xx}).

Having defined the arrival time difference ($t_{ij}$) for a pair of detectors $i$ and $j$, as well as quantified the  SN arrival time uncertainties, $\delta t$, for relevant detectors (see \cref{tab:detect}),
we adopt the following $\chi^2$ function \cite{Muhlbeier:2013gwa}
\begin{align}
\chi^2_{ij}(\alpha,\delta) = \left(\frac{t_{ij}(\alpha',\delta')-
t_{ij}(\alpha,\delta) }{\text{Max}(\delta t_i, \delta t_j)}\right)^2,
\label{eq:chi2-1}
\end{align}
 where $\alpha'$ and $\delta'$ ($\alpha$ and $\delta$) are true (tested) angular coordinates of SN and $\text{Max}(\delta t_i, \delta t_j)$ indicates that the larger $\delta t$ value is used.

 \begin{figure}
 \centering
  \begin{tabular}{cc}
    \includegraphics[width=.55\textwidth]{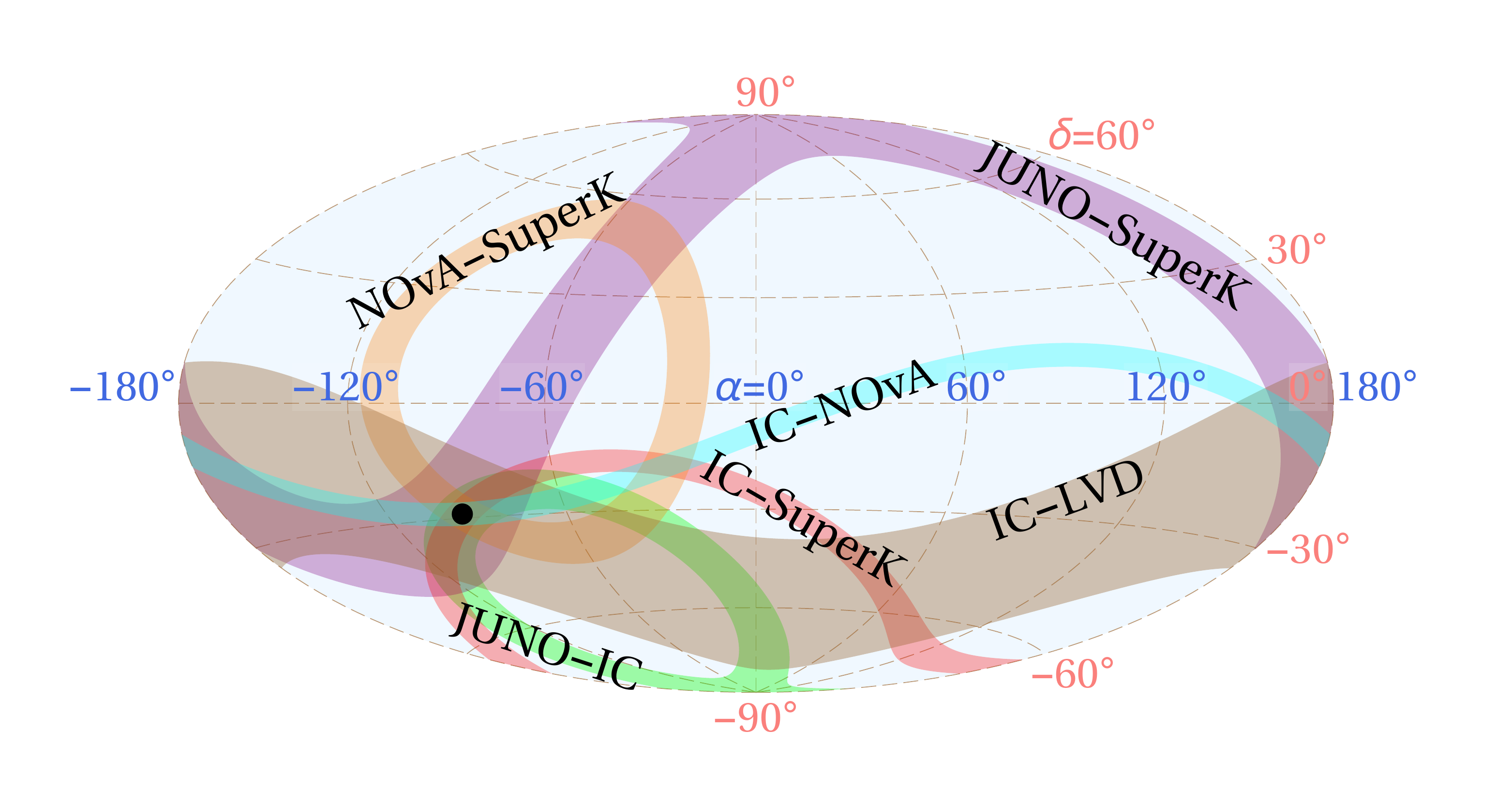} &
    \hspace{-0.05\textwidth}
    \includegraphics[width=.55\textwidth]{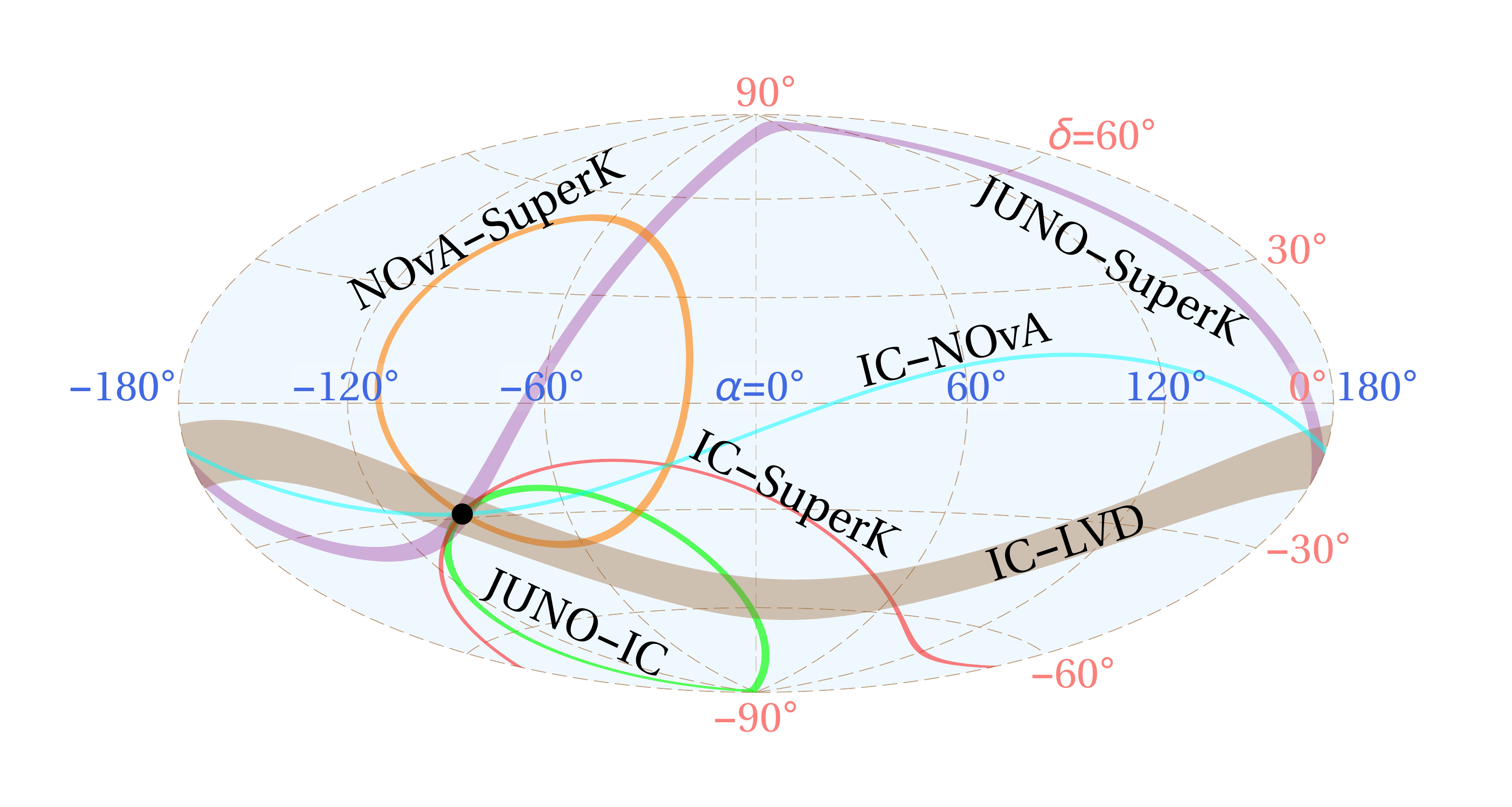} \\

  \end{tabular}
  \caption{
Regions constrained at $1\sigma$ CL by two-detector combinations.
  ``SuperK'' and ``IC'' are the abbreviations for Super-Kamiokande and Icecube detector, respectively. All regions expectedly overlap at the SN location (indicated with a black dot) which we set in the Galactic center.
 The left and right panels show scenarios of the core collapsing into a neutron star or a black hole respectively.}
  \label{fig:triangulation-2}
\end{figure}

\begin{figure}
  \centering
  \begin{tabular}{cc}
    \includegraphics[width=.55\textwidth]{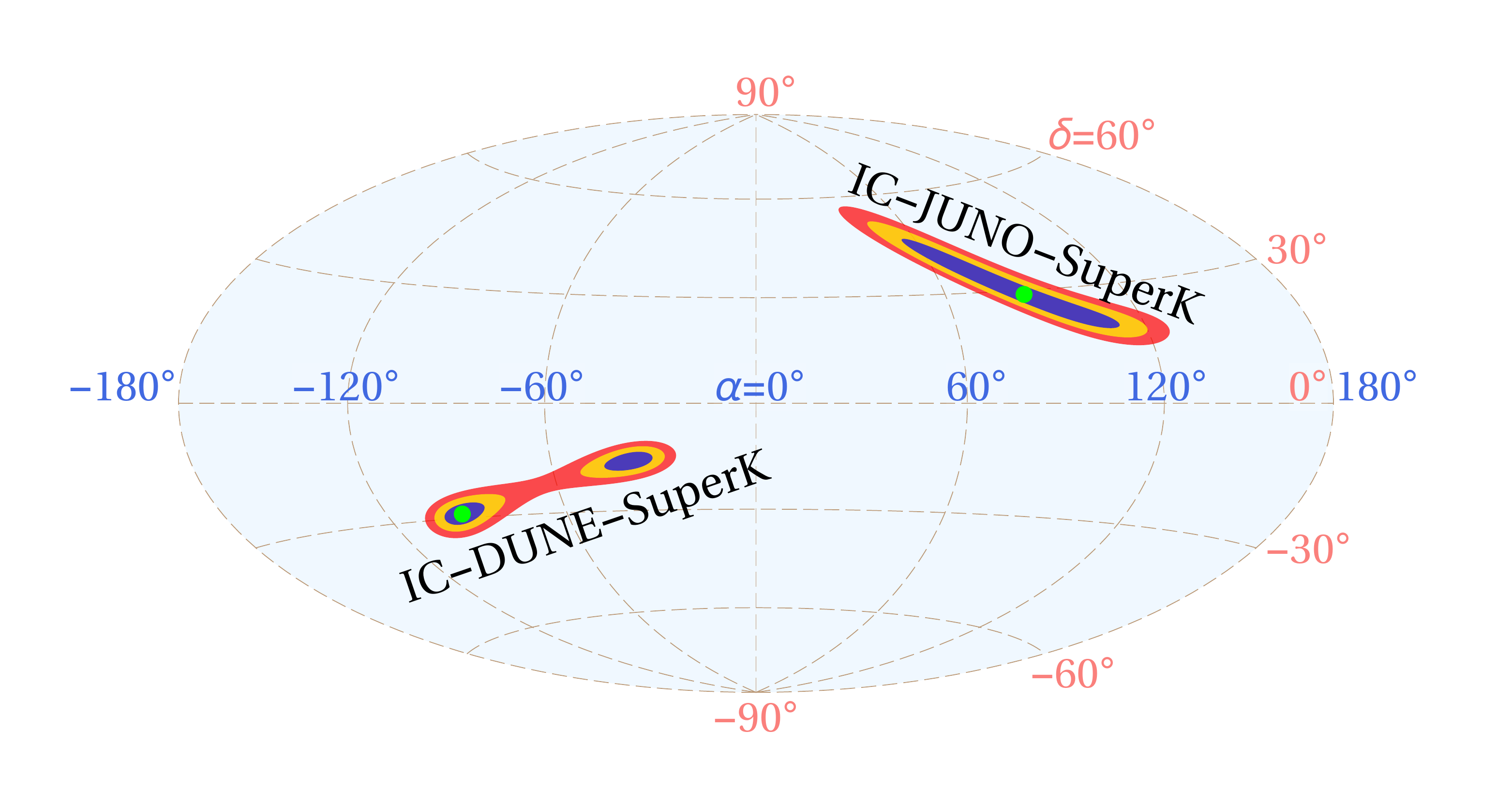} &
    \hspace{-0.05\textwidth}
    \includegraphics[width=.55\textwidth]{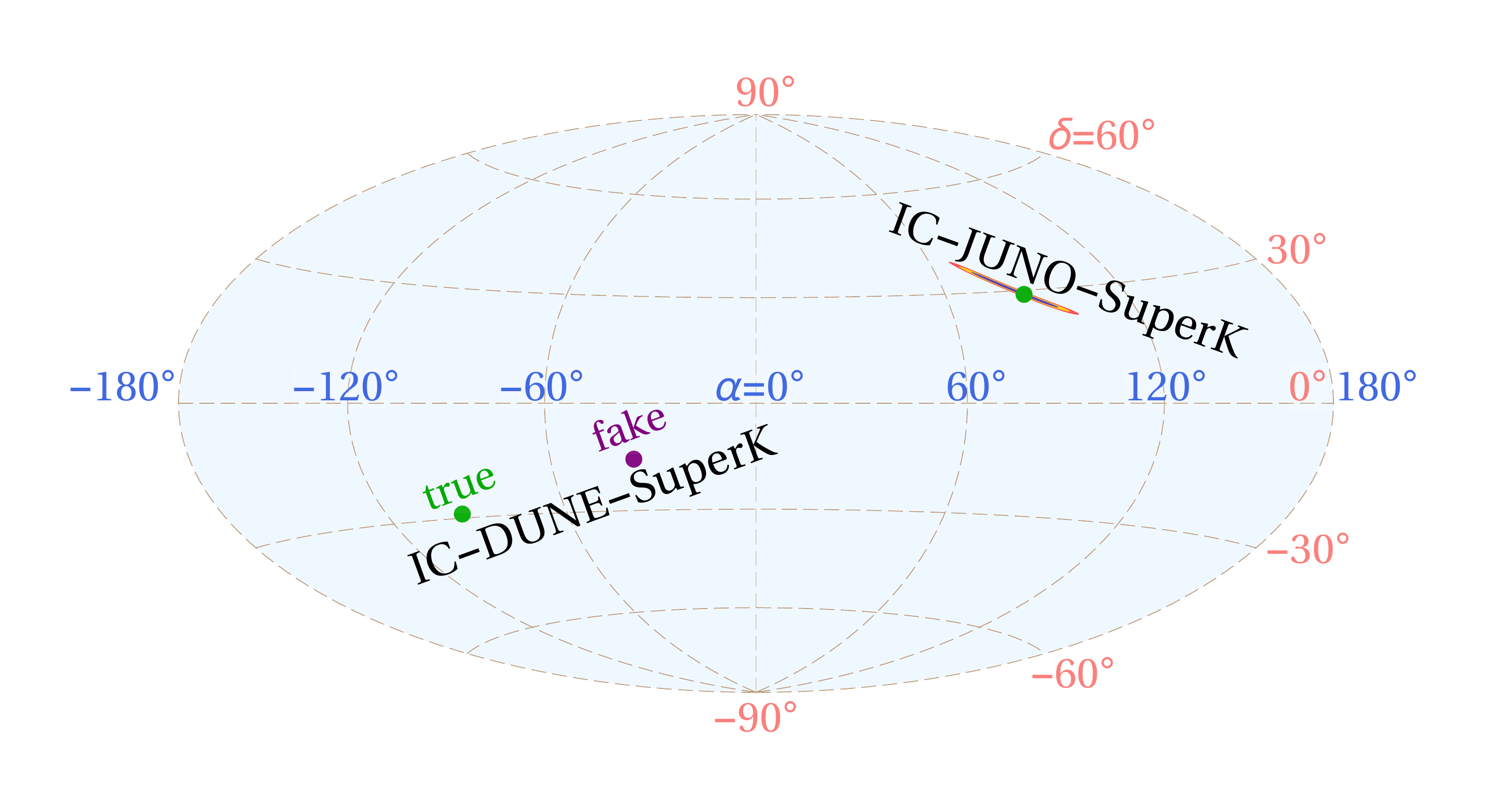} \\

  \end{tabular}
  \caption{
  In the left  panel (neutron star) we show the 1, 2 and $3\sigma$ regions for three-detector combinations.
We indicate the true SN locations with green dots. In the right panel (black hole), for the Icecube-DUNE-SuperK combination, we mark the fake solution with a purple dot. Due to high precision, the 1, 2 and $3\sigma$  regions in this case are hardly visible.
  }
  \label{fig:triangulation-3}
\end{figure}

\begin{figure}
  \centering
  \begin{tabular}{cc}
    \includegraphics[width=.55\textwidth]{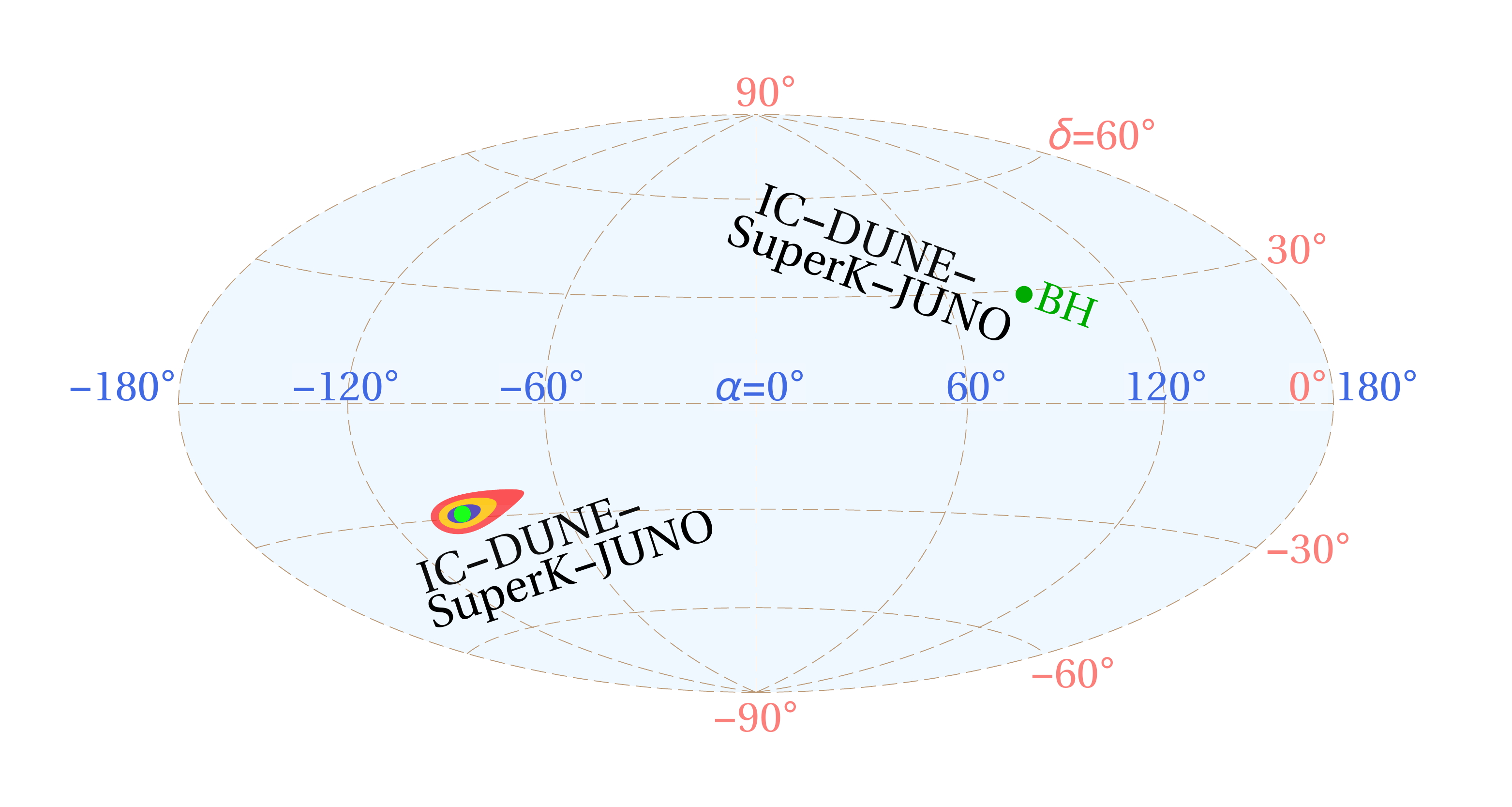} &
    \hspace{-0.05\textwidth}\includegraphics[width=.55\textwidth]{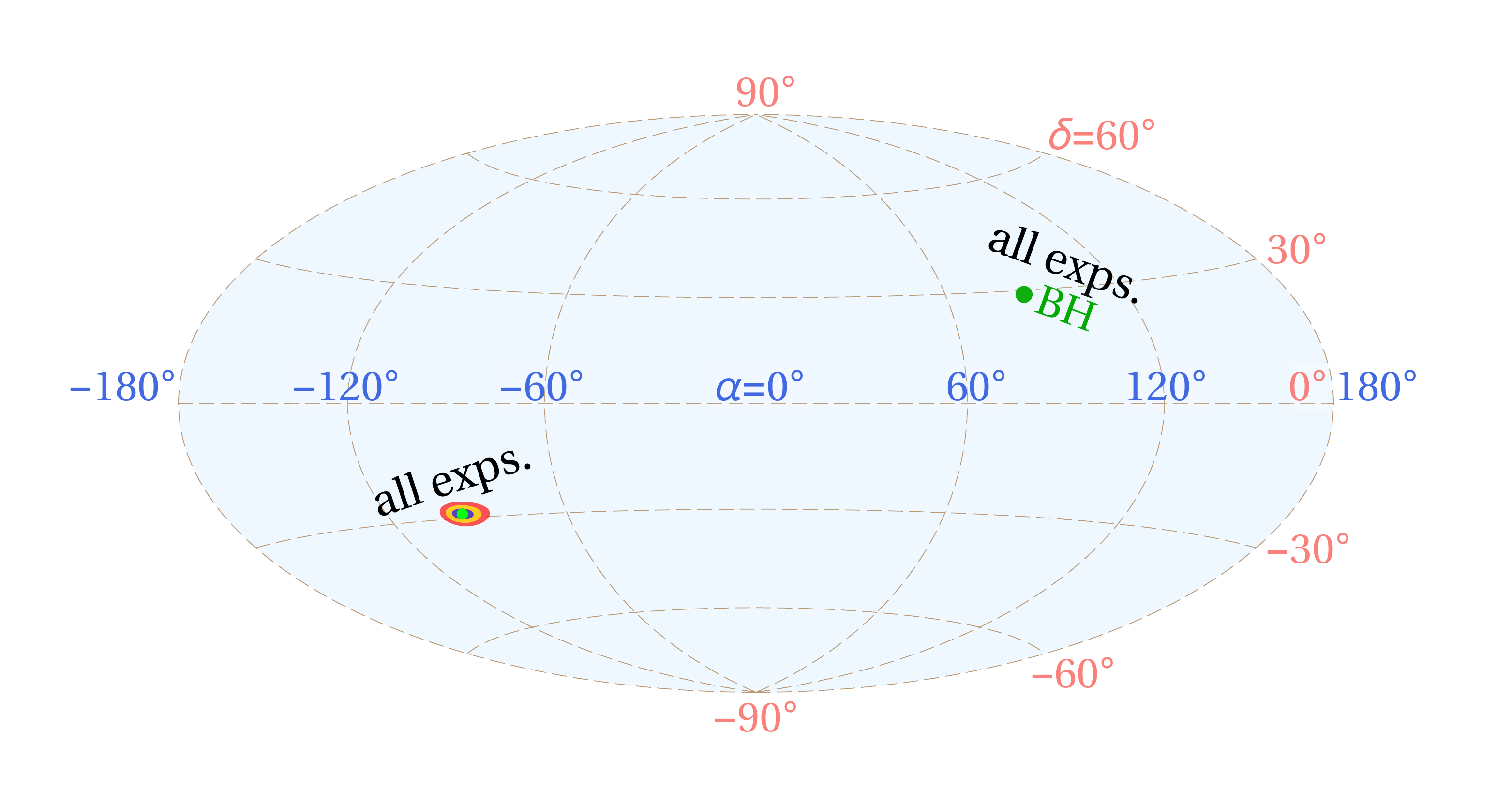} \\

  \end{tabular}
  \caption{
  In the left panel we present  a scenario  for the indicated combination of four detectors.
  The true SN location for the neutron star (black hole) scenario is set at $\alpha=-94.4^{\circ}$ and  $\delta=-28.9^{\circ}$ ($\alpha=-85.6^{\circ}$ and  $\delta=+28.9^{\circ}$).
Again,
for the black hole scenario, the 1, 2 and $3\sigma$ regions for the latter scenario are almost invisible.
In the right panel we include all considered detectors from \cref{tab:detect}.}
  \label{fig:triangulation-NS+BH}
\end{figure}

  The value of $\chi^2$  for a given $\alpha$ and $\delta$ quantifies the  likelihood that these angles coincide with the
 true ones.
 If more than two detectors are involved in the analysis, the $\chi^2$ function defined in \cref{eq:chi2-1} is summed over all detector pairs
 \begin{align}
 \chi^2_\text{tot}(\alpha,\delta)=\sum_{i,j}^{i<j} \chi^2_{ij}(\alpha,\delta).
 \label{eq:chi2-2}
 \end{align}

In \cref{fig:triangulation-2,fig:triangulation-3,fig:triangulation-NS+BH}, we
present the main results of our analysis, shown in the $\alpha$-$\delta$ parameter space with the Hammer projection. We identify the 1, 2 and $3\sigma$ regions corresponding to the statistical likelihood for the true SN position, with assumed 10 kpc radial distance from  Earth and normal neutrino mass ordering.

In \cref{fig:triangulation-2}, we show $1 \sigma$ regions for several different combinations of two detectors (indicated in the figure). The true location of SN is set to the Galactic center at $\alpha'=-94.4^\circ$ and $\delta'=-28.92^\circ$ (labeled with the black dot).
In the left (right) panel we consider the scenario of the core collapsing into a neutron star (black hole) and use the appropriate $\delta t$ values from \cref{tab:detect}. Due to the heavily improved $\delta t$ for the black hole case, the regions in the right panel are significantly smaller. Also note that, in both panels, the $1 \sigma$ regions for  JUNO-SuperK and  Icecube-LVD combinations are larger with respect to the others. For the former case,
despite detecting more than $\mathcal{O}(10^3)$ events in each detector, the
short distance between the detectors reduces the strength of the triangulation method by yielding a small numerator in \cref{eq:chi2-1}. The geographical distance between the Icecube and LVD  detectors, on the other hand, leads to a large arrival time difference ($\sim 20$ ms). However, due to the non-competitive number of events that would get detected at LVD (see \cref{tab:detect}), $\delta t$ is very large, which enhances the denominator in \cref{eq:chi2-1}. Hence, as seen in the figure, $\chi^2$ values are hard to surpass the $1 \sigma$ region in the vast portion of the parameter space.

In \cref{fig:triangulation-3}, the 1, 2 and $3\sigma$ regions are shown for the two different combinations of three detectors, namely ``Icecube-DUNE-SuperK" and ``Icecube-JUNO-SuperK". For the former combination, we assume the same SN location as in \cref{fig:triangulation-2}, whereas for the latter one we take $\alpha'=85.6^\circ$ and $\delta'=28.92^\circ$ (Galactic anticenter).
 The left (right) panel is for the core collapsing into a neutron star (black hole). As from \cref{fig:triangulation-2}, one may here again easily infer the advantage of locating SN core-collapse into a black hole. Such high precision, for one of the detector combinations, made us indicate the position of hardly visible statistical regions with a larger dot.
 What is clear from the combination of three experiments, and was already identified in Ref. \cite{Muhlbeier:2013gwa}, is the appearance of both true and fake solution. This is because the curves  $t_{ij}(\alpha,\delta)=t_{ij}(\alpha',\delta')$ in the $\alpha$-$\delta$ parameter space for  all three two-detector combinations intersect at two different coordinates. By construction (see \cref{eq:chi2-1}), these two points are then associated with $\chi^2(\alpha,\delta)=0$ value. The difference in declination and right ascension between the two solutions depends on the location of the detectors. For the ``Icecube-JUNO-SuperK" case, the region connecting the true and the fake solution is within $1 \sigma$ , whereas for ``Icecube-DUNE-SuperK" it may reach $3\sigma$ CL.

\begin{figure}
\centering
  \begin{tabular}{cc}
    \includegraphics[width=.48\textwidth]{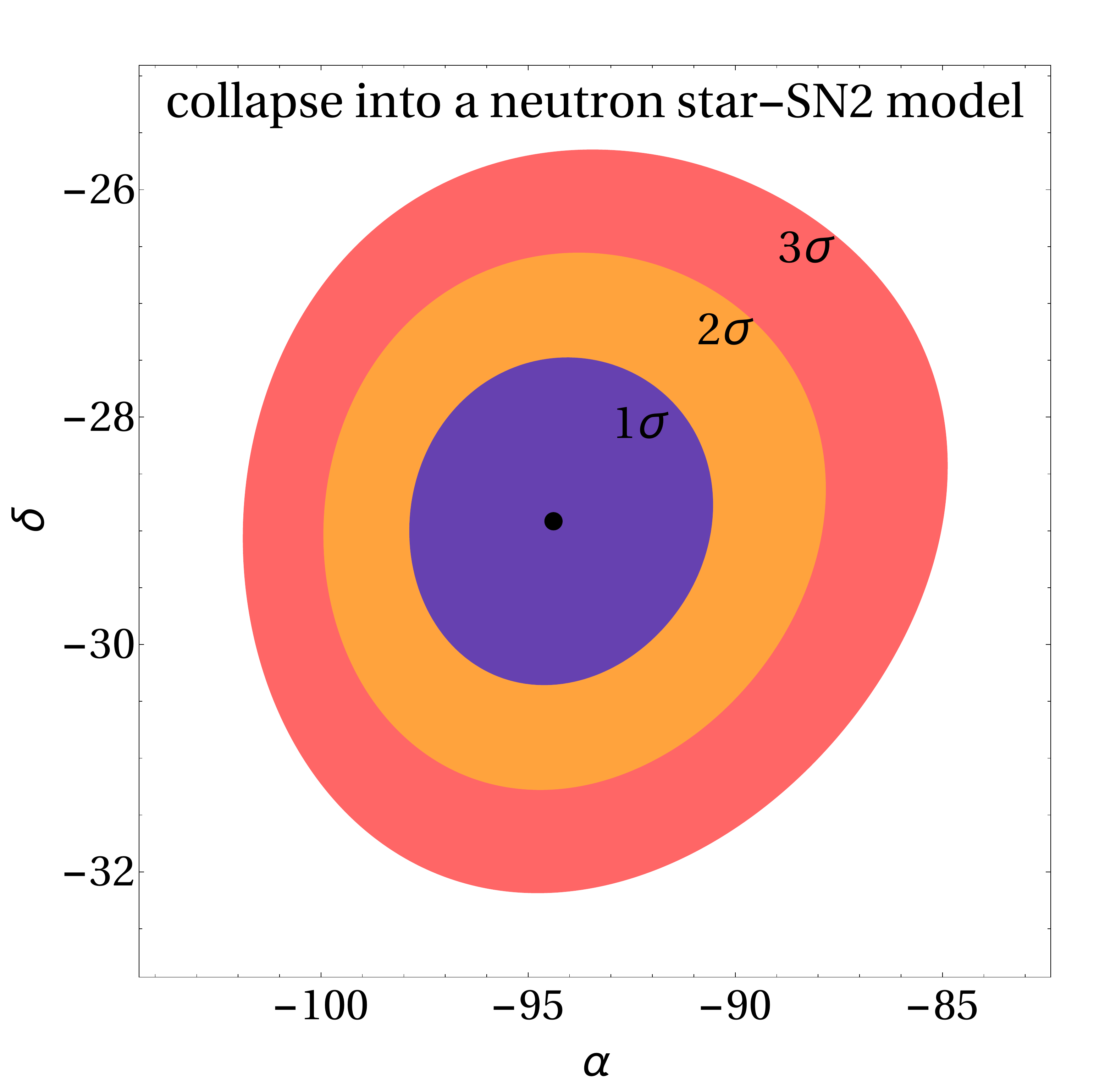} &
    \includegraphics[width=.48\textwidth]{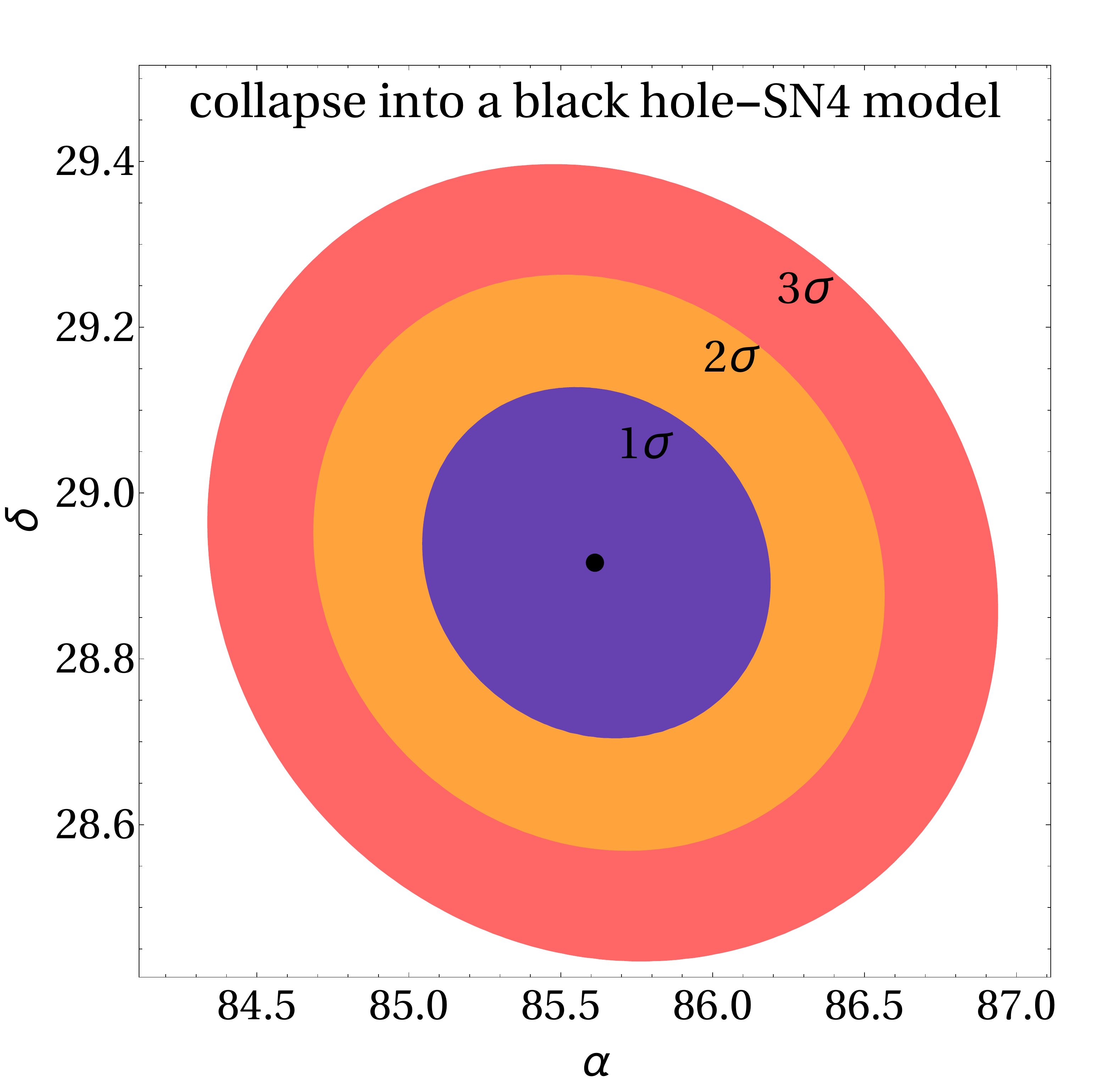} \\

  \end{tabular}
  \caption{
  Zoomed-in
  view
  of the two SN locations shown in the right panel of \cref{fig:triangulation-NS+BH}.
 The SN locations (black dots for true values) can be determined with precision of $1.5^\circ$ in $\delta$ and $3.5^\circ$ in $\alpha$ for the  neutron star case, whereas $\mathcal{O} (1^{\circ})$ precision in a black hole scenario is achieved.
  }
  \label{fig:zoom}
\end{figure}

 In \cref{fig:triangulation-NS+BH}, we show  both the neutron star and black hole scenarios, with the true SNe located in the Galactic center and anticenter, respectively. In the left panel, the combination of four indicated experiments is shown. What is readily seen is that the presence of an extra detector breaks the degeneracy of the solutions and the fake one disappears. In the right panel we use the combined power of all  considered experiments\footnote{except the DARWIN experiment which has an undetermined location.} (see \cref{tab:detect}) in order to further narrow down the significance regions. In both panels, for the same reason as in \cref{fig:triangulation-3}, the regions corresponding to the core collapsing into a black hole are indicated with larger dots.
The zoomed-in view of the right panel of \cref{fig:triangulation-NS+BH} is shown in \cref{fig:zoom}. For the core collapsing into a neutron star the width of the $1\sigma$ region is $3^\circ$ and $7^\circ$ degrees in  $\delta$ and $\alpha$, respectively. If the information on this relatively narrow range is promptly passed to the optical telescopes, there would be enough time to orient them if a favorable way and perform a complementary optical detection. For the black hole case there would be no optical signal, but the location information of the SN  would still be invaluable. One can infer from the figure, that in this case $\delta$ and $\alpha$ can be determined with sub-degree precision using the triangulation method.

\section{Summary and Conclusions}
\label{sec:summary}
In this work we studied how precisely the next Galactic supernova may be located via its neutrinos by means of the triangulation method.
For two distinct scenarios of supernova core-collapse, namely into a neutron star and a black hole,
we determined the location uncertainties in the equatorial coordinate system.
In the former scenario, precision of $1.5^\circ$  in  declination
and $3.5^\circ$ in right ascension is obtained.
For the case of the core collapsing into a black hole we demonstrated for the first time that sub-degree precision could be reached.
We envision that
this procedure may be straightforwardly  implemented and shared
 through the Supernova Early Warning System.

\begin{acknowledgments}
The authors would like to thank Evgeny Akhmedov, Stefan Br\"unner, Marta Colomer, Alec Habig, Joachim Kopp, Lutz K\"opke, Vladimir Kulikovskiy, Kai Schmitz, Zhe Wang and  Michael Wurm for very useful discussions.
\end{acknowledgments}
\bibliographystyle{JHEP}
\bibliography{refs}

\end{document}